\documentclass[aps,prd,showpacs,amsmath,nofootinbib,
superscriptaddress,twocolumn,preprintnumbers]{revtex4}
\usepackage{natbib}
\usepackage{amsmath}
\usepackage{amssymb}
\usepackage{graphicx}
\usepackage{color}
\usepackage{amsmath}
\usepackage{amsthm}
\usepackage{slashed}
\usepackage{soul}
\begin{document}

\title{Cosmology in Conformal Dilatonic Gravity}

\author{Meir Shimon}
\affiliation{School of Physics and Astronomy, 
Tel Aviv University, Tel Aviv 69978, Israel}
\email{meirs@wise.tau.ac.il}

\begin{abstract}
Gravitation is described in the context of a dilatonic theory that 
is conformally related to general relativity. 
All dimensionless ratios of fundamental dimensional 
quantities, e.g. particle masses 
and the Planck mass, as well as the relative strengths of the fundamental interactions, 
are fixed constants. An interplay between the positive energy density 
associated with relativistic matter (and possibly with negative spatial 
curvature) and the negative energy associated with dynamical dilaton 
phase results in a non-singular, flat cosmological model with no 
horizon, and -- as a direct consequence of absence of phase transitions 
in the early universe -- with no production of topological defects.
The (logarithmic) time-derivative of the field modulus is degenerate 
with the Hubble function, and all cosmological epochs of the standard 
model are unchanged except at the very early universe. We demonstrate 
that both linear order perturbation theory and the spherical collapse 
model are equivalent to those in the standard model, up to modifications 
caused by the phase of the (complex) scalar field and its perturbations. 
Consequently, our alternative theory automatically passes the main 
classical cosmological tests. Quantum excitations of the phase of 
the scalar field generate a slightly red-tilted spectrum of adiabatic 
and gaussian scalar perturbations on the largest scales. However, 
this framework does not provide a similar mechanism for producing 
primordial gravitational waves on these scales. A spherically 
symmetric vacuum solution that approximately describes the exterior 
of gravitationally bound systems (e.g., stars and galaxies) by a 
modified Schwarzschild-de Sitter metric, augmented with an additional 
linear potential term, could possibly explain galactic rotation curves 
and strong gravitational lensing with no recourse to dark matter. 
\end{abstract}

\keywords{}

\maketitle

\section{Introduction}
According to the standard cosmological model global 
evolution is driven by space expansion, namely the time-dependent Hubble scale provides 
the `clock' for the evolving properties 
of radiation and matter, resulting in a sequence of cosmological epochs. This clock is only 
meaningful if other time scales, e.g., the Planck time, or characteristic Compton times, 
evolve differently, particularly if they are 
non-varying; thus, space expansion is a relative notion.

The standard cosmological model (with an early inflationary scenario) has clearly been a very 
successful paradigm that provides a compelling interpretation 
of essentially all current cosmic microwave background 
(CMB), large scale structure (LSS) measurements, and the agreement between Big Bang 
nucleosynthesis (BBN) predictions and light element abundances.
The smallest natural length scale in the standard models of fundamental interactions, 
the Planck scale, is $O(61)$ orders of magnitude smaller than the 
Hubble scale. It is remarkable that the cosmological model provides a very good fit to 
extensive observational data, that sample phenomena over a vast dynamical range, using 
less than a dozen free parameters. 

It is this huge dynamical range that also creates one of the most vexing problems 
in theoretical physics: The stupendously delicate fine-tuning of (what 
looks like a) vacuum energy cancellation to one part in $(10^{61})^{2}=10^{122}$ 
as compared to naive expectations for the energy density of the vacuum, 
a fine-tuning known as the `cosmological constant problem'. 
Many other naturalness problems in cosmology revolve around the same 
central question of why is the universe so large and old, as compared to 
corresponding Planck scales -- `natural' scales in gravitation.

Whereas the standard cosmological model fits current measurements remarkably well, 
the essence of dark energy (DE) and cold dark matter (CDM) -- two key ingredients in the model that 
determine the background evolution, LSS formation history, and gravitational potential on galactic 
scales -- remain elusive. Additionally, 
what is considered by many as the most pristine fingerprint of cosmic inflation 
[1-3] -- a major underpinning of the standard cosmological 
model -- B-mode polarization of the CMB [4-6] induced by primordial 
gravitational waves (PGW), has not been detected. 
The latter is admittedly a very challenging measurement in the 
presence of e.g., polarized Galactic dust, nonlinear density perturbations, and 
instrumental systematics. In light of these hurdles and the allowed broad window for 
the energy scale of inflation, it is not unlikely that this signal will never be 
measured at sufficient statistical significance. Yet, its non-detection will not 
rule out the inflationary paradigm, but would rather set an (arguably very weak) 
upper bound on the energy scale of inflation.

On yet another front, the standard model (SM) of particle physics 
has also been very successful in describing the fundamental 
interactions within a relatively simple framework that rests on 
a fundamental group symmetry $SU(3)\times SU(2)\times U(1)$. 
The recent detection of the last missing building block of 
the model, the Higgs particle [7], is an impressive achievement 
of the model and provides a `proof of existence' of fundamental 
scalar fields. Indeed, at least two hypothetical key components 
of the standard cosmological model -- the inflaton field and 
quintessence -- are thought to be such scalar fields. One notable 
fine-tuning in the SM is the `hierarchy problem', i.e. the 
exceedingly small Higgs mass compared to the `natural' Planck 
scale.

The only characteristic energy scale in the SM is the vacuum 
expectation value (VEV) of the Higgs field that endows all 
elementary particles their masses. In the absence of this scale 
the SM is conformally invariant, i.e. {\it locally} scale-invariant, 
or equivalently `Weyl-symmetric'. Breakdown of the chiral symmetry 
of quantum chromodynamics (QCD) is likewise responsible for hadron 
masses. Local conformal symmetry and its breakdown are indeed vital 
for the successful working of the three known fundamental interactions 
other than gravitation.

For this and other reasons, it has long been suspected that Weyl symmetry 
(local scale invariance) also plays a fundamental role in gravitation itself. 
This old idea has been recently revived in, e.g., the suggestion that 
black hole complementarity could be potentially addressed by 
noticing that the vacuum state of a gravitational 
system is not conformally invariant [8, 9].
Various aspects of conformally symmetric extensions of the SM have been discussed in, 
e.g. [10, 11].
It has been suggested that the SM of particle physics 
on a curved background should be modified at high energies in a fashion that removes all 
scale anomalies, specifically such an amended model should result in a vanishing beta-function. 
Unfortunately, there is no unique way of achieving this ambitious 
goal that does not depend on, e.g. the choice of the scalar 
and spinor representations [12].

The main objective of the present work is to demonstrate the viability of an alternative 
physical framework based on a conformal dilatonic theory, with evolving 
fundamental dimensional `constants'. The latter are simple rescalings of 
the dilaton modulus. The theory satisfies all classical tests of general 
relativity (GR) in the solar system and on cosmological scales. 
We explore a wide range of its possible ramifications 
(albeit not exhaustively) which {\it a priori} demotes the gravitational 
constant, particle masses, and all other dimensional constants, from 
their fundamental-physical-constants status, and replaces them with 
a single dynamical (complex) scalar field (whose dynamical VEV determines 
the Planck mass). 
These VEVs have the same dynamics on the entire relevant cosmological history.
Clearly, this construction is tantamount to the Higgs 
and the dilaton VEVs being dynamical but the ratios of the proton and pion masses 
to the Planck mass is held fixed, and in addition it is equivalent to a fixed 
relative strength of the fundamental interactions.

While the intriguing possibility that the cosmological redshift could be explained 
by means of time-dependent fundamental `constants' is nearly as old as 
(what has become) the standard expanding space interpretation 
[13], it has been waived off by big bang proponents as soon as it was proposed [14]. 
This basic idea has re-emerged later within the framework of e.g., 
scalar-tensor theories of gravity [15-19] and in the context of Weyl-geometry, e.g. [20]. 
Clearly, varying `constants' could potentially have a range of cosmological 
ramifications, and since cosmological observations provide a unique window 
to peer into the remote past, cosmology is a naturally appropriate testbed 
for this possibility.

Temporal variation of the gravitational constant, $G$, is a key feature in 
scalar-tensor theories, of which Brans-Dicke (BD) 
is archetypical. Standard interpretation of observational constraints, e.g. [21], 
usually renders this theory equivalent to GR due mainly to the assumption 
that all other (dimensional) fundamental quantities are constant. 
Our approach is fundamentally different as it is based on the premise that 
{\it all} fundamental length scales have exactly the same dynamics. 
In particular, this requirement guarantees that 
{\it dimensionless} observables (in a sense that will be defined below) 
are by construction unchanged compared to their corresponding values in 
the standard cosmological model. 

Although measurement of spacetime variation of {\it dimensional} fundamental quantities, 
e.g. particle masses, $G$, $\hbar$, etc., is widely considered meaningless, 
e.g. [22-32], the literature is abound with observational constraints on the variation of 
Newton constant $G$, the speed of light $c$, Planck constant $\hbar$, 
etc. While we definitely agree with the view that only dimensionless 
ratios of dimensional quantities are unambiguous observables (in a sense 
that will be specified below), it should be emphasized that once 
fundamental dimensional quantities are promoted to dynamical fields, they do carry kinetic energy 
(and generally momentum as well) thereby affecting the dynamics of 
other fields and (what is normally understood as) the `expansion rate'. 
In addition, they may quantum mechanically (as well as thermally) fluctuate, 
thereby inducing metric perturbations (or cosmological phase transitions) in the early universe.
Such a variation of dimensional fields should only be gauged with respect to other quantities of 
the same physical units. For example, space expansion is deduced in the framework 
of the standard cosmological model and the SM -- theories that rely on the conventional system of
units where the cosmological constant characteristic scale, $l_{\Lambda}$, and the Compton 
wavelength of a massive particle, $l_{C}$, are fixed. 
In a different system of units (e.g., the `comoving frame') one may view the 
cosmological redshift as an evidence for, e.g., the monotonic contraction of the 
Compton wavelength of the emitter in a non-expanding space. 
Specifically, we will see below that within the framework of conformal 
dilatonic gravity the Hubble function is degenerate with the logarithmic time-derivative 
of particle masses and the Planck mass, i.e. with $\dot{G}/G$. 
Within the conventional system of units, null results for $\dot{G}/G$ on 
cosmological scales are therefore meaningless -- they are simply 
an `observational confirmation', merely a reflection, of our conventional 
system of units in which space is allowed to expand (i.e. the spacetime 
metric is a field) but particle masses (and the Planck mass) are fixed. 
In the comoving frame, in which energy-momentum is not conserved, the 
numerical value of $\dot{G}/G$ would be twice as large as what is normally 
considered as the `expansion rate', i.e. Hubble function $H$. In this sense, 
cosmologists have already inferred $\dot{G}/G$ averaged over cosmological scales -- 
it is $O(10^{-17})$ $sec^{-1}$ at a few percent precision (assuming the 
standard cosmological model) in the comoving frame. In contrast to $\dot{G}/G$ 
inference, the spacetime dependence/independence of the fine-structure 
of the electromagnetic force, $\alpha_{e}$ (merely a measure 
of the dimensionless ratio of the classical 
electron radius to its Compton wavelength), or of 
the dimensionless proton-to-electron mass ratio, 
is a meaningful notion since these ratios are independent of 
the field frame, and therefore do not reflect our {\it arbitrary} 
choice of field frames, i.e. of our {\it favorite} system of units, 
much like scalar functions in GR are (coordinate) frame-independent 
structures.

It is indeed conceivable in this convention that local energy-momentum 
conservation (of matter only) is no 
more than just a unit convention which implies that particle masses, 
$G$, the cosmological constant $\Lambda$, etc., are actually all fields fixed to 
their present and local VEVs. 
In particular, it is the unit convention in which gravity is described by GR, 
and the other three fundamental interactions by the SM. 
A different choice of units, i.e. rulers (in particular, dynamical rulers) would result in 
different theories of the fundamental interactions, more precisely -- 
conformally-related theories. In these theories the fundamental 
quantities, e.g. $G$, $\Lambda$, etc., do contribute to the total energy-momentum, 
and it is the generalized energy-momentum tensor, rather than the matter energy-momentum, 
which is conserved in this general case. More specifically, in GR and the SM all scalar 
fields are fixed (including the inflaton during the inflationary era) 
and all other fields 
(e.g. Dirac-, gauge- and tensor-fields) are evolving, but in a comoving frame in 
which space is static all but the scalar fields are non-evolving. In a general (field) 
frame all the fields evolve.

Generally, the `ground states' -- i.e. classical solutions derived from two conformally-related 
theories -- are not equivalent, leading in principle to falsifiable predictions.
Even the dimensionless ratio of two dynamical fields generally differs between different 
classical field configurations of the same theory, which may allow selecting the 
true vacuum, in principle.
However, data quality and parameter degeneracy often allow a range of 
possibilities as we actually argue in this work. Two 
such possibilities are the SM and GR on the one hand, 
and their `conformalized' version proposed here on the other hand. 
The fact that classical field configurations derived from two 
conformally-related theories are distinguishable, is equivalent to the statement that the 
conformal symmetry of the classical action is not a symmetry of its vacuum (i.e. ground) 
state. The underlying conformal symmetry is said to have been {\it spontaneously broken} 
in this (most general) case. 

In addition to reproducing several standard cosmological results in the present framework, 
significant insight is gained in our re-formulation of gravity. 
In this context it should be mentioned that 
the present approach also provides a convenient framework 
for alternative early universe scenarios.
Specifically, we show in the cosmological context that perturbations of 
the modulii of scalar fields which are conformally coupled to gravity 
are spurious degrees of freedom in the framework proposed here. 
Therefore, the dilaton field is promoted to a {\it complex} field; 
quantum fluctuations of its phase induce scalar metric perturbations
(unlike the modulus field, the phase is only minimally coupled to gravity). 
This is essentially the mechanism discussed in [33], that invokes a 
global $U(1)$ symmetry combined with Weyl symmetry, a mechanism that 
has no analog for producing PGW. If the latter are ultimately detected by 
upcoming or next generations CMB telescopes, via their unique 
B-mode signature [5, 6], it would certainly challenge the approach presented here. 
It should be stressed that standard cosmic inflation could be naturally accommodated by 
our framework as a solution to the relic defects, flatness and horizon problems, but not as 
a generation mechanism for primordial perturbations, as will be further discussed in section III.

The dynamical phase of the dilaton is especially 
important at small field modulii, and specifically results in a non-singular 
(classical) cosmological bounce due to, ultimately, 
the underlying global $U(1)$ symmetry.
The latter is manifested as a `centrifugal barrier' in field space; the effective 
scale factor `bounces' at its finite minimum. As mentioned above, quantum 
excitations of this phase at the very early pre-bounce phase induce scalar 
metric perturbations that survive the bounce and evolve in the post-bounce phase 
essentially as in the standard cosmological model.
This scenario naturally addresses the classical problems of the hot big bang 
model, and essentially dovetails with the standard cosmological model at the 
post-bounce cosmological phase.

On galactic (and possibly also on extra solar system) 
scales, we explore the implications of a spherically 
symmetric vacuum solution of the field equations with its non-standard spatial dependence 
on these (and smaller) scales. 
It represents deviation from standard predictions, and features 
several Weyl-gravity-like characteristics [34-37], 
largely dispensing with galactic CDM as of material origin but rather attributing 
the phenomenology of galactic CDM to a modified gravitational potential. 
This, combined with the fact that no 
alternative to CDM is offered by our model on cosmological scales, sets a lower bound 
on the possible mass range of CDM particles, rendering them possibly the most 
copiously produced particles in the early universe, and the dominant contribution 
to the entropy budget on cosmological scales.

We believe that, in addition to addressing the cosmological horizon, flatness, 
and cosmological relic problems, the framework proposed here provides 
important insight on the nature of DE, dark matter (DM), initial singularity, 
cosmological expansion, the flatness of the matter power 
spectrum on cosmological scales, the adiabatic and gaussian nature of linear 
density perturbations, and the status of conformal time 
as the fundamental coordinate parameterizing the causal 
structure of the universe on all scales (not only cosmological), rather than 
`cosmic time' $t$. The latter is simply an artifact of mass variation 
and the natural use we make of `massive clocks' in the description of 
spacetime events.  

Even so, the work presented here is by no means exhaustive, and indeed 
many of its basic aspects will be further elucidated in future papers. 
While most of the discussion is rather general we occasionally 
limit the discussion of certain aspects to specific field frames, or resort 
to particular examples and scenarios, for better communication of the key 
ideas and their implications.

Throughout this work we consider {\it only the dynamics of length scales}, 
i.e. the dynamics of scalar fields, rather than that of 
particle masses, $G$, $\hbar$, etc, to avoid unnecessary cluttering of dimensional 
`constants' and conventions that are implicitly made in, e.g. mass-to-length units 
conversions. This is a natural choice since in four-dimensional 
spacetimes scalar fields have units $length^{-1}$. In addition, lagrangian- and 
energy-densities have units $length^{-4}$, gauge- and Dirac-fields have units $length^{-1}$ 
and $length^{-3/2}$, respectively, and the {\it contravariant} spacetime metric units 
are $length^{-2}$. It is hoped 
that this systematic use of length-only units will make the 
(spontaneously broken)-scale-invariant nature of the theory, and its ramifications, more 
transparent. Our fundamental (dynamical) length scales are $l_{C}$, $l_{\Lambda}$, and 
the Planck length $l_{P}$ [Schwarzschild radius $r_{s}$ associated with a mass $M$, 
with $l_{C}=\hbar/(Mc)$, is expressed in terms of both $l_{P}$ and $l_{C}$, 
i.e. $r_{s}=O(l_{P}^{2}l_{C}^{-1})$]. 
Occasionally, we consider particle masses, the cosmological constant $\Lambda$, 
or $G$, but we do so only for clarity, when the standard parlance strongly motivates doing so. 
More generally, all fundamental length scales $l_{i}$ are set by an essentially {\it single} complex 
scalar field $\phi$ in our theory, i.e. $l_{i}=\lambda_{i}^{-1}|\phi|^{-1}$, whose time-dependent 
expectation value sets the VEV of the Higgs field, and {\it vice versa}. 
Here $\lambda_{i}$ are dimensionless Yukawa coupling parameters,  
that include the coupling parameters of the SM in addition to others, e.g. 
the dimensionless coupling of the dilaton quartic potential -- the latter is 
closely related to the cosmological constant 
and to the pseudo-conformal evolution phase posited in our early universe scenario (see section 
III.E). 
Throughout, we adopt a mostly-positive signature for the spacetime metric $(-1,1,1,1)$.

We outline our theoretical approach in section II, and the cosmological model is presented 
in section III. The spherically symmetric vacuum solution, with its ramifications on galactic 
scales, is described in section IV. 
In Section V we discuss and summarize our main results.

\section{Theoretical Framework}
A scalar-tensor theory of gravity, linear in the curvature scalar, can be 
formulated in terms of the following action given in $\hbar$ units, e.g [38, 39]
\begin{eqnarray}
\mathcal{I}&=&\int\left[\frac{1}{2}F(\phi^{K})R
-\frac{1}{2}\mathcal{G}_{IJ}(\phi^{K})\phi_{\mu}^{I}\phi^{J,\mu}
-V(\phi^{K})\right.\nonumber\\
&+&\left.\mathcal{L}_{M}(\phi^{K},\boldsymbol{\psi},{\bf A}_{\mu},g_{\mu\nu})/(\hbar c)\right]
\times\sqrt{-g}d^{4}x,
\end{eqnarray}
where the integration measure is $d^{4}x=c dt\cdot d^{3}x$, summation 
convention is implied on both greek and capital 
Latin letters, with the $N$ scalar fields $\phi^{K}$ labeled by $I, J, K=1, 2, ...., N$. 
The potential $V$ is an explicit function of the scalar 
fields. In addition to the spacetime metric $g_{\mu\nu}$ 
we introduce $\mathcal{G}_{IJ}$, merely a sigma-model-type metric in field space. 
$R$ is the curvature scalar calculated from $g_{\mu\nu}$ and its first 
and second derivatives in the usual way, and 
$f_{\mu}\equiv f_{,\mu}\equiv\frac{\partial f}{\partial x^{\mu}}$.
The matter Lagrangian, $\mathcal{L}_{M}$, which is allowed to explicitly depend on 
$\phi^{K}$, as well as on other fields such as, Dirac ($\boldsymbol{\psi}$), 
metric ($g_{\mu\nu}$), and gauge ($\boldsymbol{A}_{\mu}$) fields, 
accounts for the entire contribution to the energy density budget of 
the universe (i.e. DE, DM, baryons, electrons, neutrinos, and radiation) 
except for the kinetic energy associated with the scalar fields, 
and their non-minimal coupling to gravitation.
Although Eq. (1) emphasizes the role of $V(\phi^{K})$ -- a self-interaction 
term associated with purely the dilaton field -- the latter could be 
equally well absorbed in a redefined $\mathcal{L}_{M}(\phi^{K})$.

Throughout, we work in a basis in which $\mathcal{G}_{IJ}$ is diagonal. 
Depending on the sign of its coefficients the kinetic term 
$-\frac{1}{2}\mathcal{G}_{IJ}(\phi^{K})\phi_{\mu}^{I}\phi^{J,\mu}$ can be either 
non-negative or non-positive. 
In the cosmological context, if the kinetic energy associated with one of the 
scalar fields is negative then its effective perfect fluid equation of state (EOS) 
is generally $w\lesssim -1$ [40]. The corresponding perfect fluid description appropriate 
for an underlying field (either scalar or Dirac) theory can be obtained by, e.g. incoherently 
averaging over field modes [41]. Recent inference of the Hubble function from 
observations at the local universe suggests that space might be expanding faster than 
was previously deduced based on CMB observations. These observations [42, 43] could 
be explained, among other viable possibilities, by a 
`phantom'-like behavior of cosmic acceleration driven by slowly varying energy density 
of a scalar field with $w\lesssim -1$ [44]. We focus in this work on only one such a field 
with negative kinetic energy, which we identify as simply the cosmic scale factor.
It is plausible that one (or more) of the scalar fields of Eq. (1) with a 
negative kinetic term is responsible for the apparent recent cosmic acceleration.

The field equations derived from variation of Eq. (1) with respect 
to $g_{\mu\nu}$, and $\phi^{K}$ are, respectively, the 
generalized Einstein equations, the scalar field equations, and the 
generalized energy momentum (non-) conservation, e.g. [38, 39]
\begin{eqnarray}
F\cdot G_{\mu}^{\nu}&=&T_{M,\mu}^{\nu}+\Theta_{\mu}^{\nu}-\delta_{\mu}^{\nu}V\\
\Box\phi^{I}&+&\frac{1}{2}F^{I}R-V^{I}
+\Gamma^{I}_{JK}\phi^{J\alpha}\phi^{K}_{\alpha}=-\mathcal{L}_{M}^{;I}\\
T_{M,\mu;\nu}^{\nu}&=&\mathcal{L}_{M,J}\phi_{\mu}^{J}.
\end{eqnarray}
The effective energy-momentum tensor associated with the scalar fields, $\Theta_{\mu}^{\nu}$, 
and the `connection' $\Gamma^{I}_{JK}$ constructed from the field metric, are 
\begin{eqnarray}
\Theta_{\mu}^{\nu}&\equiv&\mathcal{G}_{IJ}\left(\phi^{I}_{\mu}\phi^{J,\nu}
-\frac{1}{2}\delta_{\mu}^{\nu}\phi^{I,\alpha}\phi^{J}_{\alpha}\right)
+F_{\mu}^{\nu}-\delta_{\mu}^{\nu}\Box F\\
\Gamma^{I}_{JK}&\equiv&\frac{1}{2}\mathcal{G}^{IL}(\mathcal{G}_{LJ,K}
+\mathcal{G}_{LK,J}-\mathcal{G}_{JK,L}). 
\end{eqnarray}
Here and throughout, $f_{\mu}^{\nu}\equiv(f_{,\mu})^{;\nu}$, with $f_{;\mu}$ 
denoting covariant derivatives of $f$, $\Box f$ is the covariant Laplacian, 
$(T_{M})_{\mu\nu}\equiv\frac{2}{\sqrt{-g}}\frac{\delta(\sqrt{-g}\mathcal{L}_{M})}
{\delta g^{\mu\nu}}$, and $H_{I}\equiv\frac{\partial H}{\partial\phi^{I}}$ 
for any function $H$. Latin indices are raised and lowered by 
means of $\mathcal{G}^{IJ}$ and $\mathcal{G}_{IJ}$, respectively.
Eq. (4), which is not independent of (2) \& (3), 
implies that energy-momentum (of matter alone) is generally not 
conserved, which is expected in the case that $G$, $\Lambda$, or particle masses are 
spacetime-dependent. In particular, we 
specialize Eq. (1) to a theory conformally-related to GR. Under a 
conformal transformation of the spacetime metric $\tilde{g}_{\mu\nu}=\Omega^{2}(x)g_{\mu\nu}$, 
where $\Omega(x)$ is an arbitrary spacetime-dependent function, local energy-momentum 
conservation $T_{M,\mu;\nu}^{\nu}=0$ 
is replaced by $\tilde{T}_{M,\mu;\nu}^{\nu}=-(\Omega_{,\mu}/\Omega)\tilde{T}_{M}$, where 
$\tilde{T}_{M}$ is the trace of the energy-momentum tensor. 
For Eq. (1) to be conformally-related to GR scalar fields have to similarly transform under 
conformal transformations (as shown below), 
and particle masses become dynamical, thus violating energy-momentum conservation. 
In the special case of photon gas, or any other species 
characterized by a traceless energy-momentum tensor, energy-momentum 
conservation is frame-independent. Consequently, massless particles still follow geodesics 
in this theory but massive particles do not. We will argue in the following sections that this 
{\it is} the origin of cosmological redshift and non-Keplerian behavior observed in 
galactic rotation curves -- phenomena that are normally attributed to space expansion 
and galactic DM, respectively. 

Assuming $\mathcal{G}_{IJ}$ are independent of the fields in a `cartesian' basis in 
field space (i.e. all scalar fields have length units $length^{-1}$), and that 
$F\equiv -\zeta|\phi|^{2}$ where $|\phi|^{2}=-\sum_{IJ}\mathcal{G}_{IJ}\phi^{I}\phi^{J}$ with constant $\zeta$, and 
combining Eq. (3) with the trace of Eq. (2), we obtain the following consistency relation
\begin{eqnarray}
&&\mathcal{G}_{IJ}(1+6\zeta)\left(\phi^{I}\Box\phi^{J}+\phi^{I,\mu}\phi^{J}_{,\mu}\right)\nonumber\\
&=&T_{M}-\rho_{M,I}\phi^{I}+V_{I}\phi^{I}-4V.
\end{eqnarray}
Of particular interest -- and the focus of this work -- 
is the case $\zeta=-1/6$ for which (the generally dynamical) 
Eq. (7) significantly simplifies and reduces to a constraint 
equation. In this case Eq. (7) integrates to $\rho_{M}\propto|\phi|^{1-3w_{M}}$ 
where $w_{M}$ is the matter EOS parameter, and $V$ has been absorbed in $\mathcal{L}_{M}$ 
with an effective EOS parameter $w=-1$. 
As expected, $\rho_{M}$ is a quartic potential in the case $w_{M}=-1$, and is independent of 
$|\phi|$ in the case $w_{M}=1/3$.

It can be shown that with $F=-\zeta\phi^{2}$ the {\it vacuum} 
of our fundamental action in Eq. (1) 
is equivalent to a BD theory with $\omega_{BD}=(4\zeta)^{-1}$ (where $\omega_{BD}$ is 
the BD dimensionless parameter). For a general $F$, which is not necessarily quadratic in the scalar field, 
the generalized coupling parameter $\omega_{BD}(\phi)=-F/(F')^{2}$ 
(assuming the $\mathcal{G}_{IJ}$ metric components are fixed constants) becomes $\phi$-dependent.
Our choice, $\zeta=-1/6$, corresponds to conformal coupling of the scalar field 
to gravity, which is equivalent to $\omega_{BD}=-3/2$, e.g. [45]. 
In the original BD proposal [46] the matter lagrangian $\mathcal{L}_{M}$ 
does not explicitly depend 
on the scalar field. Whereas BD represents a genuine departure from GR by, e.g. allowing 
for a spacetime-dependent relative strength of gravity and the other interactions, our construction 
below guarantees that the relative strengths of the fundamental interactions are fixed, 
as mentioned above.
This is achieved by a non-trivial transformation of the matter lagrangian between the frames, 
and in this sense our theory, with non-vanishing matter contributions, 
is significantly different from that of Brans \& Dicke with $\omega_{BD}=-3/2$ [45]. Notably, BD 
required $\omega_{BD}>0$ to guarantee the positivity of the Hamiltonian in their original 
proposal [46], a fact that was emphasized and reinterpreted in [47]. The instability of BD 
theories with $\omega_{BD}<0$ was further emphasized in 
[48], but we will see, as mentioned above, that non-positive kinetic terms are not an issue in our 
theoretical framework. This discussion has been resurrected in many forms (e.g. [49] and references 
within). The `anomalous' nature of this particular BD theory has 
been pointed out also in other, though related, contexts, e.g., [50, 51]. Below 
we argue that in a field frame in which cosmological evolution 
is carried entirely by scalar rather than metric 
dynamics, the kinetic term associated with this scalar field must 
be negative when considered as a contribution to the total energy budget.
The observational lower limit $\omega_{BD}\geq 40000$ reported in [21] is 
synonymous to Eq. (1) (with $F=\zeta\mathcal{G}_{IJ}\phi^{I}\phi^{J}$)
essentially reducing to GR (with a reversed relative sign between the gravitational and the 
matter parts of the action if one insists on a non-negative kinetic term for the dilaton). 
As explained above, this conclusion would 
be true only if all other dimensional quantities, such as particle masses and the cosmological 
constant, are indeed fixed constants (vanishing in particular, 
but not necessarily) as was originally proposed by 
BD, which is not the case with our model, Eq. (1). 

The Einstein-Hilbert (EH) action 
with a positive cosmological constant $\Lambda$, is 
$\mathcal{I}_{EH}=(2\kappa\hbar)^{-1}\int (R-2\Lambda+\mathcal{L}_{M})\sqrt{-g}d^{4}x$, 
where $\kappa\equiv 8\pi G/c^{4}$.
The specific choice $\zeta=-1/6$ in Eq. (1) corresponds to GR in 
dynamic units, with $\phi^{K}$ conformally coupled 
to gravitation, i.e. it is the `conformal', or equivalently `Weyl-symmetric', 
version of the EH action. In this case Eq. (5) is the generalization 
of the `improved' (traceless) energy-momentum tensor [45, 52] to the case of multiple 
scalar fields. Any conformal rescaling of the metric, scalar, Dirac and gauge fields, 
as $g_{\mu\nu}\rightarrow\Omega^{2}(x)g_{\mu\nu}$ 
(i.e. $g^{\mu\nu}\rightarrow\Omega^{-2}(x)g^{\mu\nu}$), $\phi\rightarrow\Omega^{-1}(x)\phi$, 
$\psi\rightarrow\Omega^{-3/2}(x)\psi$, and $A_{\mu}\rightarrow\Omega^{-1}(x)A_{\mu}$, 
respectively [with an overall $\mathcal{L}_{M}\rightarrow\mathcal{L}_{M}\Omega^{-4}(x)$], 
brings Eq. (1) (with $\zeta=-1/6$ and $V=\lambda|\phi|^{4}$ with a constant $\lambda$) 
to its `Einstein frame' (EF), i.e. $\mathcal{I}_{EH}$, form. 
Therefore, dimensionless polynomial ratios of fields are invariant to this field 
re-definition, but ratios involving derivatives thereof, e.g. kinetic terms, are 
no longer invariant under conformal transformations.
We emphasize that a symmetry of $\mathcal{I}_{EH}$ 
is not necessarily a symmetry of the classical field configurations derived from it. 
As mentioned above, in the case that the underlying conformal symmetry is not 
respected by the classical solutions conformal symmetry is said to have 
been `spontaneously broken'. This property allows us to obtain new classical solutions, 
non-trivially related to the corresponding GR solutions, as is illustrated 
with a specific example in section IV. 

If the structure of the SM could provide 
any guiding principles for the fundamental nature of gravity it 
would probably be its {\it local} gauge invariance, as well as 
its near {\it local} scale-invariance (with the VEV of the Higgs field being 
the only dimensional scale in the SM).  
Therefore, we focus on the case 
$F\equiv -\frac{1}{6}\sum_{IJ}\mathcal{G}_{IJ}\phi^{I}\phi^{J}$ 
in spite of the triviality of the `Weyl-current' in this case, 
e.g. [51, 53, 54]. 
A few alternative cosmological models based on {\it global} scale 
invariance -- not conformally-related to GR -- have been explored 
in e.g., [15-19, 53, 54]. 

Our assumption is that the combination of a conformal version 
of SM and GR is the correct classical representation of the four known fundamental 
interactions, e.g. [10]. In fact, conformal dilatonic gravity extended to the SM is power-counting 
renormalizable as it contains no dimensional constants, and should in principle have a 
fixed UV point of the renormalization group. Classical GR coupled to the SM could represent 
just one vacuum state out of infinitely many possible vacuua.
The underlying conformal symmetry is {\it spontaneously broken} by this ground state 
which is characterized by constant VEV of the Higgs and dilaton fields, that are manifested 
by constant particle masses and a fixed Planck scale, respectively. This {\it ad hoc} 
choice may not necessarily align with the ground state actually realized in nature. 
Astrophysical and cosmological observations may have already revealed that the actual `vacuum 
state' is different from our conventional choice, as alluded to above and further discussed 
in sections III and IV below.

In subsequent sections it is shown that, at least in the cosmological context,
perturbations of real scalar fields are spurious degrees of freedom in the proposed framework -- 
they can be absorbed in redefined metric and density perturbations. 
In the standard cosmological model density and scalar metric 
perturbations are sourced by the inflaton field fluctuations. Therefore, this 
mechanism for generating perturbations 
cannot be readily invoked in the framework advocated here. Instead, we consider 
a complex scalar field. In particular, its phase perturbations are genuine degrees of freedom 
that seed primordial scalar perturbations. In addition, the resulting cosmological model 
is devoid of the initial singularity due to the phase dynamics at small modulii values 
governed by a global $U(1)$ symmetry (see section III.E).
The gravitational sector of our complex field model applied to Eq. (1) is
\begin{eqnarray}
\mathcal{I}&=&\int\left[\frac{1}{6}|\phi|^{2}R+\phi^{*}_{\mu}\phi^{\mu}-\lambda|\phi|^{4}
+\mathcal{L}_{M}(|\phi|)/(\hbar c)\right]\nonumber\\
&\times&\sqrt{-g}d^{4}x.
\end{eqnarray}
The Newtonian limit of the theory described 
by Eq. (8) is obtained in section III.D. The quartic potential 
form is required by conformal invariance (dimensionless 
coupling $\lambda$) in a scalar-tensor theory. However, as we saw below 
Eq. (7), other polynomial degrees are allowed in $\rho_{M}$ 
in the presence of Dirac and gauge fields, while still maintaining conformal invariance.
A natural amendment to Eq. (8) might be the addition of the term 
$\mathcal{I}_{W}=-\alpha_{W}\int C_{\alpha\beta\gamma\delta}C^{\alpha\beta\gamma\delta}\sqrt{-g}d^{4}x$, 
essentially fourth-order Weyl gravity. Here, $C_{\alpha\beta\gamma\delta}$ is the Weyl tensor 
and $\alpha_{W}$ is a dimensionless coupling constant. 
Possible relation between the two theories on a more fundamental level has been 
discussed in, e.g. [55-58]. Weyl gravity has its own merits [58], a few of them 
might be relevant in the broader context of a combined conformal dilatonic gravity 
and Weyl gravity theory. However, for the rest of this work we assume that 
gravitational dynamics on macroscopic scales is governed by conformal dilatonic 
gravity alone, Eq. (8).
Embedding the SM of particle physics in the theory described by Eq. (8) 
is straightforward [59]. However, this goal is not pursued in 
the present treatment. Instead, as a conformally-related alternative to GR 
the proposed model Eq. (8) automatically guarantees that the ratio of particle masses 
to the Planck mass is fixed. 
Introducing a (conformally-coupled and an effectively) dynamical Higgs VEV 
implies that no primordial phase transitions had occurred, and no relic 
topological defects have been thus generated. It should be emphasized that 
the recent detection [7] relies on reactions involving the Higgs coupling 
to quarks and gauge bosons, but not its self-interactions. Thus, the existence 
of a new scalar field has been confirmed but the underlying purported spontaneous 
symmetry breaking mechanism that crucially depends on the `Higgs potential' 
has not been corroborated yet.

While the underlying conformal symmetry of the action Eq. (8) is 
spontaneously broken, this fundamental symmetry is restored in a few special cases. 
Note the difference between the action Eq. (8) being invariant under 
the canonical dimension scaling, $\phi\rightarrow \Omega^{-1}$, $\psi\rightarrow \psi\Omega^{-3/2}$, 
$A_{\mu}\rightarrow A_{\mu}\Omega^{-1}$, and $g_{\mu\nu}\rightarrow g_{\mu\nu}\Omega^{2}$, 
and the (stronger) requirement that a given solution 
of the field equations is $\phi\propto\sigma^{-1}(x)$, $\psi\propto\sigma^{-3/2}(x)$, 
$A_{\mu}\propto\sigma^{-1}(x)$, and $g_{\mu\nu}\propto\sigma^{2}(x)$, where $\sigma(x)$ 
is an arbitrary spacetime function. 
The latter case would represent a `conformal' solution to the field equations that 
does not result in any spacetime evolution. Such a solution seems to be (cosmologically) 
non-viable as it does not describe an evolving cosmological model 
as seems to be clearly required, e.g., 
by observations of metallicity evolution, cluster abundance evolution, 
and CMB (dimensionless) 
temperature evolution with redshift. 
We encounter this limiting `symmetry restoration' behavior at very early/late epochs 
({\it large} negative/positive conformal times) in the cosmological context (section III) and at very large 
distances in our spherically symmetric static vacuum solution in section IV. 

It is the conventional expectation that conformal symmetry is restored at high energies, 
but these two examples may imply that Weyl symmetry is a symmetry of the 
ground-state on large length scales 
rather than high energies.
If all the fields do scale at some regime by their canonical 
dimensions then the dynamical field equations reduce to 
algebraic equations that constrain combinations of 
the integration constants -- basically the VEVs of these fields -- 
as we see in one specific example in section III.C. 
In the cosmological context, this regime of the theory 
might be relevant to a few well-known `cosmological coincidences'.

We emphasize that only dimensionless ratios of polynomials are invariant 
in our framework, not ratios involving kinetic terms 
(except for regimes where conformal symmetry has been 
restored, as was described above). Generally, dimensionless ratios involving both 
kinetic and potential terms do evolve. The fact that the Hubble 
time scale (clearly associated with a kinematic rather than polynomial term in the 
field equations) generally evolves with respect to other fundamental time scales, 
e.g. the Planck time or Compton interaction times 
(which are monomials in the scalar fields), thereby representing 
a cosmologically evolving universe, is a manifestation of the spontaneously broken conformal symmetry 
(except for purely cosmological constant domination eras). 

For the rest of the paper, `dimensionless ratios' will always refer to dimensionless 
{\it polynomial} ratios. While the latter are required to be fixed 
only to the precision level of current observations, this has other advantages beyond 
aesthetics. First, it reduces the problem to that of essentially a single complex scalar 
field conformally coupled to gravity from which the dynamics of all other fundamental length scales is 
readily obtained via rescaling by fixed dimensionless constants, i.e. ratios of Yukawa couplings. 
Second, all dimensionless coupling constants are globally spacetime-independent, 
implying that the relative strengths of the various fundamental interactions 
do not change in transforming between frames since Eq. (8) is frame-independent. 
In particular, the dimensionless couplings regulating the strengths of all four fundamental interactions are 
fixed, by construction, consistent with astrophysical and cosmological probes of the cosmic evolution of 
at least the fine structure constant $\alpha_{e}\equiv e^{2}/(\hbar c)$, e.g. [60], with our understanding of BBN, and other early universe processes. 
Current constraints on fine structure variation from BBN, the CMB, quasar spectra, etc., show no evidence for 
variation of these dimensionless coupling constants. 
Analogously, the dimensionless coupling constant regulating 
the gravitational interaction between two masses $m_{1}$ \& $m_{2}$, 
$\alpha_{g}\equiv G m_{1}m_{2}/(\hbar c)$, is $\propto m_{1}m_{2}/M_{Pl}^{2}$, 
where $M_{Pl}$ is the Planck mass. Since all masses in our construction scale 
proportionally to the scalar field it then follows that $\alpha_{g}=constant.$

While the Higgs mechanism is thought to endow elementary particles their masses, composite 
particles, e.g. pions and protons, are believed to obtain their masses via chiral 
symmetry breaking of quark condensates. The VEV of these condensates $\langle\bar{\psi}\psi\rangle$, 
normally considered as a dimensional number of units $length^{-3}$, 
is here promoted to a cubic scalar term $\propto\phi^{3}$. This scaling guarantees that, e.g., 
the proton-to-electron mass ratio is fixed, to avoid conflict with observations. 
This VEV, exactly like the VEV of the Higgs field 
(essentially proportional to the rescaled dilaton) 
discussed above, effectively conformally couples to gravity 
(via a $\frac{1}{6}R|\phi|^{2}$ term), and therefore has the appropriate 
spacetime evolution that guarantees all dimensionless mass ratios are 
spacetime-independent. 

The respective spacetime metrics obtained from GR and our conformal dilatonic gravity are either 
identical (up to conformal rescaling), or deviate only at extreme conditions, 
not directly probed by astrophysical 
observations, or are interpreted differently. For example, the standard interpretation of galactic CDM 
is replaced within the Weyl gravity framework 
(as well as by the corresponding solution within conformal dilatonic gravity, see section IV 
below) by modified Newtonian dynamics at large distances from the galactic center [34]. 

Conformal dilatonic gravity should not be confused with fourth-order Weyl conformal 
theory of gravity [34-37, 58, 61-64]. While both theories are 
conformally symmetric, the former includes a dynamical scalar field absent 
from the latter. Whereas the field equations associated with Weyl gravity are fourth order, 
the equations describing conformal dilatonic gravity are only second order. It should be mentioned that 
in the non-vacuum case matter is conformally coupled to gravity within fourth-order Weyl gravity in 
exactly the same fashion as considered in the present work within the conformal dilatonic theory of 
gravitation. 
However, since this prescription has been employed to matter fields only, the kinetic terms 
associated with them are non-negative and they contribute negatively to the effective $G$ [62]. In 
contrast, we propose below that the modulii of all scalar fields are conformally coupled to gravity. 
If, in addition to the dilaton, another scalar field that is supposed 
to describe matter is conformally coupled to gravity, it must then contribute negatively to the 
effective $G$ in Eq. (8), and positively to the kinetic term.  
Overall, $G>0$ and gravity is attractive due to the huge hierarchy between the Planck 
and other hypothetical scalar fields. 
Finally, the Weyl term quadratic in the Weyl tensor is absent from Eq. (1), so even if 
the prescription of matter coupling to gravity of [62] is extended to scalar field with negative kinetic 
terms, the theories still differ for non-conformally-flat spacetimes. 

\section{Cosmological Model}
We describe the background evolution, the evolution of linear perturbations, 
the spherical collapse model, and a singularity-free early universe scenario, characterized 
by scale-free, gaussian and adiabatic density perturbations but with no analogous PGW production.

\subsection{Redshift in Comoving Frame}
Before we discuss the proposed model in more detail, it is of interest to 
point out a fundamental difficulty with the standard interpretation of cosmological redshift, 
a phenomenon usually attributed to momentum decay of photons in an adiabatically expanding space. 
However, this argument does not readily hold in comoving coordinates: Making 
the coordinate transformation $dt\rightarrow a d\eta$, where $a(\eta)$ and $\eta$ 
are the scale factor and conformal time, respectively, photons effectively travel 
in a static space, the energy density of radiation is fixed, and no redshift 
would occur due to space expansion. 
Massive particles that follow timelike geodesics still `see' an expanding space.
Of course, redshift is now absorbed in the newly-defined $\eta$. However, 
we could have equally well started with comoving rather than cosmic coordinates, 
pretending we know nothing about `cosmic time', i.e. with an 
infinitesimal line element $ds^{2}=a^{2}[-d\eta^{2}+\frac{dr^{2}}{1-Kr^{2}}
+r^{2}(d\theta^{2}+\sin^{2}\theta d\varphi^{2})]$, which implies that in terms 
of these comoving coordinates (using conformal rather than cosmic time) photons and 
massive particles `see' essentially two different metrics differing by a 
multiplicative factor $a^{2}(\eta)$. Photon momentum clearly does not decay in this 
coordinate system and cosmological redshift can only be explained by particle masses 
scaling as $a(\eta)$. Indeed the governing lagrangian of a massive 
point particle is $\mathcal{L}_{pp}=-mc\int\frac{ds}{d\mu}d\mu$ 
(where $\mu$ parameterizes the geodesic followed by the particle) 
and $a(\eta)$ could be absorbed in a redefined mass $m\rightarrow ma$, 
explicitly breaking energy conservation in a homogeneous and isotropic universe. We then 
arrive at the paradoxical situation that the only explanation in the standard 
cosmological model for the observed cosmic redshift in the comoving frame entails 
the breakdown of a central underpinning of the underlying theory, GR.
In the following sections we explore the merits of describing the cosmological model in an 
arbitrary field frame, as opposed to the standard EF. 

We emphasize the fact that just making the transformation 
$m\rightarrow ma$ in the comoving frame is insufficient to capture the 
essence of varying masses for at least two reasons. First, this procedure ignores the energy 
and momentum associated with this dynamical mass at the action level, potentially 
ignoring an important contribution to the total energy budget. 
Second, this `mass' may quantum-mechanically or thermally fluctuate. The former is 
usually associated with seeding primordial density perturbations, as in inflation. The 
latter normally drives cosmological phase transitions.

While we keep the discussion as general as possible, 
and usually do not commit to particular field frames, we occasionally focus on a special 
`Jordan frame' (JF), `the comoving frame', i.e. a frame at which space is static, 
to illustrate basic ideas. Specifically, in the comoving frame only scalar fields 
evolve, and all other fields, including the spacetime metric, have constant amplitudes.
As mentioned above, the energy density associated with the CMB does not evolve in 
the comoving frame. Note that this does not conflict with past 
inference of temperature `evolution' of the CMB according to the adiabatic 
scaling $T_{CMB}\propto 1+z$ from observations of molecular transitions 
toward distant clouds or the Sunyaev-Zeldovich (SZ) effect towards 
galaxy clusters, because these measurements, (indeed, any other such measurement) 
are only sensitive to dimensionless quantities involving $T_{CMB}$, not to $T_{CMB}$ itself. 
For example, molecular line transitions measure the ratio of photon 
energy to the rotational energy spectrum of the molecule, e.g. [65-67]. 
The latter is only determined by 
their masses and therefore this measurement is sensitive to the evolution of the photon 
wavelength relative to the effective length associated with the molecular structure. 
Similarly, measurements of $T_{CMB}(z)$ towards galaxy clusters, e.g. [68-70], 
are only sensitive to the dimensionless frequency 
of the CMB, $h\nu/(kT(z))$, not to $T(z)$ alone. While these measurements certainly exclude, e.g.  
steady-state cosmologies in which these dimensionless quantities are constant, they are blind 
to conformal transformations of dimensionless polynomial ratios [e.g. $h\nu/(kT(z))$], 
i.e. a cosmological model in the comoving frame passes these tests 
exactly as does the standard cosmological model. 

\subsection{Temporal Evolution of Length Scales}
The discussion in this section is confined to the comoving frame to allow 
better transparency of the main features; many of the results are generalized 
in the following sections to an arbitrary field frame. 
Here, we essentially 
reformulate the EH action as a conformally coupled dilaton theory defined 
on a static spacetime background.

Consider the vacuum EH action with a positive cosmological constant $\Lambda$, 
$\mathcal{I}_{EH}=(2\kappa\hbar)^{-1}\int (R-2\Lambda)\sqrt{-g}d^{4}x$, where $\kappa\equiv 8\pi G/c^{4}$. 
The Friedmann-Robertson-Walker (FRW) action 
$\mathcal{I}_{FRW}=(\kappa\hbar/3)^{-1}\int(-a'^{2}+Kc^{2}a^{2}-\Lambda a^{4}/3)\sqrt{-\tilde{g}}d^{4}x$ 
is obtained by taking $R=6(a''/a+Kc^{2})/a^{2}$ and 
integration by parts; $\tilde{g}$ is the static metric conformally related to the FRW metric, 
$\tilde{g}_{\mu\nu}\equiv g_{\mu\nu}/a^{2}=diag(-1,\frac{1}{1-Kr^{2}},r^{2},r^{2}\sin^{2}\theta)$, 
with $K$ being the spatial curvature, a prime denotes derivatives with 
respect to conformal time, and the time coordinate in the volume element $d^{4}x$ is conformal. 
One can readily verify that the Euler-Lagrange equation for $a$ that extremizes the action $\mathcal{I}_{FRW}$ 
is indeed the vacuum Friedmann equation $\mathcal{H}^{2}+Kc^{2}=\frac{\Lambda a^{2}}{3}$, 
where $\mathcal{H}\equiv a'/a$ is the conformal Hubble function. 
The kinetic term associated with the scale factor $a$ in 
$\mathcal{I}_{FRW}$ is always non-positive, and the effective potential $-Kc^{2}a^{2}+\Lambda a^{4}/3$ 
is monotonically non-decreasing for $K\leq 0$ \& $\Lambda\geq 0$.

Defining the modulus of a complex scalar field $\phi=\rho e^{i\theta}$ as 
$\rho^{2}\equiv 3a^{2}/(\kappa\hbar c)$, the FRW action is reformulated as a
conformally symmetric scalar field theory (now with matter accounted for 
by the lagrangian $\tilde{\mathcal{L}}_{M}$) with a non-positive kinetic term, 
defined on a static background 
\begin{eqnarray}
\mathcal{I}_{FRW}&=&\int\left(\partial_{\mu}\phi\partial^{\mu}\phi^{*}-V(\phi)
+\tilde{\mathcal{L}}_{M}(|\phi|)\right)\nonumber\\
&\times&\sqrt{-\tilde{g}}d^{4}x.
\end{eqnarray}
Here, $V(\phi)\equiv -K|\phi|^{2}+\lambda|\phi|^{4}$ is an effective potential, 
$K<0$ and $\lambda\equiv(8\pi\Omega_{\Lambda}H_{0}^{2}t_{Pl}^{2})/3>0$, with 
$\Omega_{\Lambda}$, $H_{0}$ and $t_{Pl}$ denoting the energy density associated 
with the cosmic term in critical density units at present, the Hubble parameter, 
and the Planck time, respectively. The 4D infinitesimal volume element is 
$\sqrt{-\tilde{g}}d^{4}x=\sinh^{2}\chi\sin\theta d\eta d\chi d\theta d\varphi$, 
where $\chi$ is a `radial' coordinate in the hyperbolic coordinate system.
Comparison of Eq. (9) with Eq. (8) implies that the curvature scalar 
associated with comoving space is $R=6K$. This is expected since our model is 
equivalent to a conformal transformation of the metric 
$g_{\mu\nu}\rightarrow\Omega^{2}g_{\mu\nu}$ with $\Omega=a^{-1}$ 
in the specific case of the comoving frame, then the curvature 
scalar becomes $R=6K$. It is fixed and nonsingular; the curvature 
singularity has been absorbed by the scalar field $\phi$. 
Since the curvature scalar transforms inhomogeneously, i.e. 
$R\rightarrow\Omega^{-2}(R-6\Box\Omega/\Omega)$, under the transformation 
$g_{\mu\nu}\rightarrow\Omega^{2}g_{\mu\nu}$, curvature singularities can be 
lifted and replaced by scalar field singularities, i.e. those of $\Omega$. 
From the perspective advocated in this work the initial cosmological 
{\it curvature singularity} is a frame-dependent singularity.
The two degrees of freedom of the complex scalar 
field are more than sufficient to rectify the (scalar) curvature 
singularity, but are generally insufficient for lifting Riemann tensor singularities 
describing the curvature of lower symmetry spacetimes.  
We emphasize that the fundamental actions considered in this work is Eq. (8). 
In contrast, Eq. (9) was basically recovered from the Friedmann equation 
for illustrational purposes; $K$ is an integration constant of the classical 
field equations rather that a dimensional parameter appearing in the conformally-invariant 
action, Eqs. (8). By analogy, we argue that the Higgs VEV is fundamentally 
dynamical, and it is only due to our experience with the spontaneously broken symmetry 
in a given field frame (see discussion in section IV) that we are prompted to plug 
a fixed Higgs VEV in the conventional SM action.

The monotonic potential $V(\phi)\equiv -K|\phi|^{2}+\lambda|\phi|^{4}$ 
(with $K<0$ \& $\lambda>0$) 
implies that there can be no static field configuration describing the ground state; 
`space' (i.e. $|\phi|$) must be either expanding or contracting. 
As mentioned in the preceding section, 
in this static metric frame radiation does not redshift and likewise gauge 
fields only oscillate with constant amplitudes. 
Similarly, the number density of non-relativistic (NR) fermions, 
$\propto\bar{\psi}\psi$, is fixed in this homogeneous 
and static space and therefore the amplitude of $\psi$ is fixed as well. 
Consequently, only the scalar fields evolve in this frame, replacing the standard metric 
(i.e. the scale factor) evolution 
of the FRW spacetime. The effective matter lagrangian in Eq. (9) is
$\tilde{\mathcal{L}}_{M}\equiv(\kappa/3)^2/(\hbar c)\mathcal{L}_{M}-C_{\theta}^{2}/\rho^{2}$, 
where we made use of the fact that the global $U(1)$ symmetry of the scalar field $\phi$ 
in Eq. (8) implies that $\rho^{2}\theta'=C_{\theta}$, 
and $C_{\theta}$ is an integration constant. Thus, our choice of a complex 
(rather than real) 
scalar field effectively induces a new (negative) contribution to the total energy density. 
This contribution functions as a `centrifugal barrier' at small $|\phi|$ values  
and ultimately results in a nonsingular cosmological model as is shown in section III.E. 

Concordance model values for $\Omega_{\Lambda}$ and $H_{0}$ imply that $\lambda=O(10^{-122})$, 
indeed an unnaturally small dimensionless number which clearly reflects the fine tuning 
of the cosmic energy density (dominated by the `cosmological constant' term) 
to $\approx 122$ decimal places in Planck energy density units 
(note that the degree of fine-tuning is exactly the same as the one 
familiar from the standard cosmological model since our potential is quartic).
However, there is a significant distinction between these numerically identical fine-tuning 
levels in our formalism and in the standard cosmological model. While naive estimates 
in the latter for the energy density associated with the vacuum, 
$\rho_{vac}=\int^{E_{Pl}} E^{3}dE=O(M_{Pl}^{4})$, are outrageously 
{\it discrepant} with the observationally-inferred $\Omega_{\Lambda}$ by $\approx 122$ orders 
of magnitude, the present approach does not make any such prediction 
since $\Omega_{\lambda}$ here, like $\Omega_{DM}$ in the standard cosmological model, 
are all free model parameters. In fact, as a {\it parameter}, the Planck 
mass $M_{Pl}=E_{Pl}/c^{2}$ (or for that matter the GUT or the electroweak scales)
does not even exist in our reformulation of the FRW action, and therefore 
cannot serve as a UV momentum cutoff. Likewise, the hierarchy problem associated with the Higgs 
mass is associated with this `natural cutoff' normally set on radiative corrections. 
The problem of replacing the external mass cutoff in standard renormalization 
with a scale-invariant regularization scheme is not yet settled, but a would-be 
scale-invariant procedure is expected to invoke the (dynamical) VEVs of the scalar 
fields rather than imposing arbitrary external dimensional parameters that explicitly 
break the underlying conformal symmetry, e.g. [54, 71]. 

Admittedly, explaining the stupendously small value of 
the dimensionless $\lambda$ remains an open problem in our framework, 
indeed a severe fine-tuning. 
However, this fine-tuning 
is a direct result of assuming that $\Lambda$ appearing in the EH action, 
$\mathcal{I}_{EH}=(2\kappa\hbar)^{-1}\int(R-2\Lambda)\sqrt{-g}d^{4}x$, is strictly a 
constant, and of $\lambda|\phi|^{4}$ in Eq. (9) being observationally very small 
in comparison to the Planck energy density $O(|\phi|^{4})$. Alternatively, had we 
replaced $-\lambda|\phi|^{4}$ in Eq. (9), 
with $\partial_{\mu}\chi\partial^{\mu}\chi^{*}+(K/\sqrt{\lambda})|\chi|^{2}-\tilde{\lambda}|\chi|^{4}$, 
where $\chi$ is a new field that satisfies $\chi\sim\lambda^{1/4}\phi$, then we 
would have obtained that $\tilde{\lambda}=O(1)$. 
Moreover, the kinetic term associated with $\chi$, i.e. $\partial_{\mu}\chi\partial^{\mu}\chi^{*}$, 
is dwarfed by the corresponding dilaton kinetic term, $\partial_{\mu}\phi\partial^{\mu}\phi^{*}$, 
over the entire cosmic history. In other words, at sufficiently 
small $\phi$, i.e. insofar the quartic terms in Eq. (9) are subdominant, then 
$\phi$ and $\chi$ have exactly the same dynamics and introducing $\chi$ 
merely amounts to a redefinition of $\phi$. When the cosmic 
term $\Lambda=\tilde{\lambda}|\chi|^{4}$ starts dominating it would behave as would 
$\lambda|\phi|^{4}$ but without the enormous (122 orders of magnitude) fine-tuning. 
Instead, this still leaves us with a huge hierarchy ($\sim 30$ orders of magnitude) between the $\phi$ 
and $\chi$ VEVs. We recall that some $\sim 17$  
orders of magnitude hierarchy already exists between the electroweak and GR theories.

\subsection{Evolution of the Cosmological Background}
In this section the background evolution is described in a (field) frame-independent 
fashion. In particular, the degeneracy between the FRW scale factor 
and the modulus of our scalar field, essentially particle masses, 
is highlighted. As concrete examples, we discuss several 
aspects of the cosmological model in the standard frame and one particular JF, 
the comoving frame. Other important features of the cosmological model 
are discussed in subsequent sections.

The Einstein tensor components $G_{\mu}^{\nu}$, associated with 
the metric $g_{\mu\nu}=a^{2}\cdot diag(-1,\frac{1}{1-Kr^{2}},r^{2},r^{2}\sin^{2}\theta)$,
with conformal rather than cosmic time, are 
\begin{eqnarray}
G_{\eta}^{\eta}&=&-3a^{-2}\left(\mathcal{H}^{2}+Kc^{2}\right)\nonumber\\
G_{i}^{j}&=&-a^{-2}\left(2\mathcal{H}'+\mathcal{H}^{2}+Kc^{2}\right)\delta_{i}^{j},
\end{eqnarray}
where we used the conformal Hubble function $\mathcal{H}\equiv a'/a$.
Note that $f'\equiv\frac{\partial f}{\partial\eta}$ is the derivative of 
a function $f$ with respect to conformal time $\eta$.
Here, `$i,j$' indices stand for the spatial coordinates. 

Although the focus of this work is the special case $\zeta=-1/6$, 
it is constructive to consider the general case first. 
With $\mathcal{G}_{IJ}=-2(1,1)$, $\phi\equiv\rho e^{i\theta}$ in polar field coordinates, 
$F=-2\zeta\rho^{2}$, and $V=-\lambda\rho^{4+\alpha}$ (with $\alpha$ a dimensionless constant), 
applied to Eqs. (2), we obtain the generalized Friedmann equation (Eq. A1), a linear 
combination of this equation with the generalized Raychaudhuri equation (Eqs. A1 \& A2), 
and the second-order equation governing the modulus field dynamics (Eq. A3), respectively
\begin{eqnarray}
&&\mathcal{F}^{2}-6\zeta\mathcal{H}^{2}-12\zeta\mathcal{H}\mathcal{F}-6\zeta K\nonumber\\
&=&\tilde{a}^{2}\tilde{\rho}_{M}+\lambda\tilde{a}^{2}\rho^{\alpha}-\theta'^{2}\\
&-&6\zeta\tilde{\mathcal{H}}'-6\zeta\mathcal{H}^{2}-(1+12\zeta)\mathcal{F}^{2}-12\zeta\mathcal{H}\mathcal{F}
-6\zeta K\nonumber\\
&=&\frac{(1-3w_{M})}{2}\tilde{a}^{2}\tilde{\rho}_{M}+2\lambda\tilde{a}^{2}\rho^{\alpha}+\theta'^{2}\\
&-&6\zeta\mathcal{H}^{2}+\mathcal{F}^{2}+2\mathcal{H}\mathcal{F}-6\zeta\mathcal{H}'+\mathcal{F}'-6\zeta K\nonumber\\
&=&\frac{(1-3w_{M})}{2}\tilde{a}^{2}\tilde{\rho}_{M}+(2+\alpha/2)\lambda\tilde{a}^{2}\rho^{\alpha}+\theta'^{2},
\end{eqnarray}
where the energy-momentum tensor of a perfect fluid is 
$(T_{M})_{\mu}^{\nu}=\rho_{M}\cdot diag(-1,w_{M},w_{M},w_{M})$, 
and we defined the generalized scale factor $\tilde{a}\equiv a\rho$,  
$\mathcal{F}\equiv\rho'/\rho$, and the energy 
density scales as $\tilde{\rho}_{M}=\rho_{M_{0}}\tilde{a}^{-3(1+w_{M})}$. 
Note that the EOS of perfect fluids does not change under conformal 
transformations, i.e. $\tilde{w}_{M}=w_{M}\equiv P_{M}/\rho_{M}$.
Eq. (13) can then be replaced by its combination with Eq. (12)
\begin{eqnarray}
-(1+6\zeta)(\mathcal{F}'+2\mathcal{F}^{2}+2\mathcal{H}\mathcal{F})
=-\frac{\alpha}{2}\lambda\tilde{a}^{2}\rho^{\alpha}.
\end{eqnarray}
The equations considerably simplify once $\zeta=-1/6$ is fixed. First, the consistency relation, Eq. (15), 
implies that the potential must be quartic, as we saw earlier in Eq. (7) for general spacetime metrics. 
Second, defining $\tilde{\mathcal{H}}\equiv\tilde{a}'/\tilde{a}$, Eqs. (11) \& (12) considerably simplify 
and can be written is a manifestly frame-invariant fashion as follows
\begin{eqnarray}
\tilde{\mathcal{H}}^{2}+K&=&\tilde{a}^{2}\tilde{\rho}_{M}+\lambda\tilde{a}^{2}-\theta'^{2}\\
\tilde{\mathcal{H}}'+\tilde{\mathcal{H}}^{2}+K&=&\frac{(1-3w_{M})}{2}\tilde{a}^{2}\tilde{\rho}_{M}
+2\lambda\tilde{a}^{2}+\theta'^{2},
\end{eqnarray}
i.e. the generalized Friedmann equations are manifestly invariant under any simultaneous 
change of $\rho$ and $a$ that leaves their product, $\tilde{a}$, unchanged. 

Frame invariance of Eqs. (15) \& (16) 
reflects the fact that the ground state of Eq. (8) is degenerate. The standard FRW spacetime 
with constant particle masses, $G$ \& $\Lambda$, is just one possible ground state ($\rho=constant$). 
It should be emphasized, however, that in practice this degeneracy should leave no 
fingerprints on the classical motion of test particles in this particular case 
(we will encounter a counter example in section IV). As mentioned in section III.A, null 
geodesics are blind to conformal transformations, and timelike geodesics are derived 
from the action $\mathcal{L}=\frac{m}{\hbar c}\int\frac{ds}{d\tau}d\tau$, 
which is proportional to $a\rho$ and is therefore invariant 
to conformal transformations. It is also clear from 
comparison of these equations with the classical Friedmann equation that the conserved 
quantity is $\rho_{M_{0}}\tilde{a}^{-3(1+w_{M})}$ rather than simply $\rho_{M_{0}}a^{-3(1+w_{M})}$. 
As expected this $\rho-a$ degeneracy carries over to linear perturbation theory 
as shown in section III.D. 

The Machian nature of gravity -- arguably, a virtue 
of BD theories -- is most apparent in this model once the 
choice $a=1$, i.e. $\mathcal{H}=0$, is made: The modulus $\rho$ (i.e. particle masses, $G^{-1/2}$) 
is ``chameleonic'' in the sense that it follows the total energy density [the $\theta'^{2}$ term is absorbed 
as a `stiff matter' component in a redefined $\rho_{M}(\rho)$].  
Actually, our model Eq. (9) fits better with generalized BD theories (more so than that of BD) 
where the matter lagrangian explicitly depends on the dilaton field, 
(as in Bergmann-Wagoner gravity [72, 73]), i.e. both $G$ and particle masses are spacetime-dependent.

As can be readily read off from Eqs. (15) \& (16) 
the coupling constant $\lambda$ plays the role of the cosmological constant given in Planck units, 
i.e. $O(l_{P}^{2}\Lambda)$. These equations are supplemented by 
two more field equations obtained from variation of the action with respect 
to the scalar fields, as in Eq. (3). It is easily verified that the equation 
for the modulus $\rho$ obtained from Eq. (A3) is 
redundant with Eq. (15). Whereas $\tilde{a}=a\rho$ essentially replaces the 
scale factor $a$ of the FRW model, $\theta$ plays no role in the standard 
cosmological model. As already mentioned and as shown in 
section III.E below, inclusion of a non-vanishing $\theta'$ in our cosmological model 
serves two purposes. First, it could be used to avoid the initial singularity. 
Second, it couples excitations of $\theta$ to metric perturbations, i.e. 
to perturbations of the gravitational potential as discussed in section III.E 
below. 
Now, to close our system of equations, the evolution equation of 
the phase, obtained using Eq. (A3), is,
\begin{eqnarray}
\theta''+2\tilde{\mathcal{H}}\theta'=0.
\end{eqnarray}

The comoving frame is characterized by a static background $a=1$ \& $\mathcal{H}=0$,  
in contrast to the standard cosmological model that features $\rho=\rho_{0}$ \& 
$\mathcal{F}=0$ where $\mathcal{F}\equiv\rho'/\rho$ (and essentially $\theta=0$). 
If we ignore for a moment the $\theta'$ terms it is clear 
from Eqs. (15) \& (16) that the static space model is obtained from that in the standard cosmological model  
by simply making the replacement $a\rightarrow\rho/\rho_{0}$ and 
$\rho_{M}\rightarrow\rho_{M}(\rho/\rho_{0})^{4}$, where $\rho_{0}$ is the 
present value of $|\phi|$, which relates to Newton constant as $\rho_{0}^{2}=O(G^{-1})$.
This implies, as we saw earlier in section II, 
that $\rho_{M}\propto\rho^{1-3w_{M}}$ since in the standard cosmological model 
$\rho_{M}\propto a^{-3(1+w_{M})}$. Indeed, $\rho_{M}\rightarrow|\phi|^{4}\rho_{M}$ as 
we transform from the standard cosmological model to the JF and that 
$a^{-1}\propto l_{c}\propto|\phi|^{-1}$, i.e. $|\phi|\propto a$, or 
more precisely; $\rho/\rho_{0}=a$ where we set $a_{0}=1$.

Now, assume that in the comoving frame $\theta'\neq 0$, then Eq. (17) implies 
that $\theta'\propto\rho^{-2}$, and by virtue of Eqs. (15) or (16) 
this additional contribution could be considered an effective `stiff matter` 
contribution with $w_{\theta}=1$ that would dominate the cosmic evolution 
at early epochs, i.e. in the small field limit $\rho\rightarrow 0$. A few authors entertained 
the possibility that the early cosmic evolution in the standard expanding cosmology was at one point 
dominated by an effective 
stiff matter with $\rho_{M}=P_{M}$, e.g. [74, 75]. This scenario has interesting 
implications for PGW production, the CMB, etc. However, since BBN leaves only a 
narrow wiggle room around $w=1/3$, e.g. [76], such a term must have 
decayed by the time BBN commenced. 
We explore the impact of this effective stiff energy density contribution to the total 
budget on the background evolution and linear perturbations in section III.E.

As mentioned, the amplitude of a Dirac field associated with fundamental 
particles, such as the electron, does not decay over time in the comoving 
frame because the number density of NR fermionic particles, 
assuming the evolution is adiabatic, is unchanged. 
However, quarks, the building blocks of hadrons are described by Dirac fields 
and their masses are replaced by the more fundamental scalar fields; quark masses 
evolve exactly as the VEVs of the Higgs dilaton fields do.
Likewise, fermion condensates do evolve in our theoretical framework. 
Therefore, the global cosmological evolution is encapsulated in the evolving scalar 
fields, i.e. fundamental length scales, e.g. Planck length scale and Compton wavelengths.

Not surprisingly, and much like in the EF, 
Eq. (15) is easily found to be the first integral of Eq. (16).
This leaves us with Eq. (15) \& (17) [and with $\rho_{M}\propto\tilde{a}^{-3(1+w_{M})}$, 
i.e. the generalized energy-momentum conservation, Eq. (4)] 
as the only independent equations that determine 
the background dynamics. By comparison to the FRW model in the case of constant EOS, 
the proportionality constant is easily recognized to be $\propto n_{M}^{1+w_{M}}$, 
where $n_{M}$ is the number density 
of matter particles (which could be relativistic or not), resulting in
\begin{eqnarray}
\rho_{M}/(\hbar c)\propto\rho^{1-3w_{M}}n_{M}^{1+w_{M}}.
\end{eqnarray}
This relation is easily verified to be the correct scaling in a few {\it constant} 
values of the EOS parameter; 
dust ($w=0$), radiation ($w=1/3$), and DE ($w=-1$). 
Indeed, the energy density of NR matter is linear in 
masses, and the energy density of radiation is independent of masses. 
For example, in the case of NR fermions $n_{M}\propto\psi\bar{\psi}$, where $\psi$ is 
the Dirac field with fixed amplitude. In the comoving frame 
only scalar fields evolve, and both gauge and fermionic fields are fixed in the static 
background -- thereby providing a typical length scale, e.g. $[\psi]=length^{-3/2}$.
In other words, the fields do not generally scale by their canonical dimensions.

The energy-momentum tensor of matter itself is known to be generally non-conserved 
in scalar-tensor theories of gravity, as is evident from Eq. (4). 
By virtue of the formal equivalence to the Friedmann equation 
in the standard cosmological model the conserved 
quantity in the homogeneous and isotropic background is $\rho_{M}/\rho^{4}$. Another way 
to see this is to consider the continuity equation, which in the standard cosmological model 
reads $\rho'_{M}+3(1+w_{M})\mathcal{H}\rho_{M}=0$. 
The analogous (non-) conservation equation in a static background model is Eq. (4), 
$\rho'_{M}=\rho_{M,\rho}\rho'$, then $\rho'_{M}=\mathcal{F}\rho_{M,\rho}\rho$. Substituting this 
relation in $\left(\rho_{M}/\rho^{4}\right)'+3(1+w_{M})\mathcal{F}\left(\rho_{M}/\rho^{4}\right)=0$ 
the correct scaling $\rho_{M}\propto\rho^{1-3w_{M}}$, Eq. (18), is recovered.

Since space does not expand in the comoving frame, 
the only evolution of the NR energy density is due to evolving 
masses. The Friedmann equation results in $\rho\propto\eta^{2/(1+3w_{M})}$ 
(in epochs during which the effective single-fluid EOS varies weakly) 
which is a monotonically increasing function for any $w_{M}>-1/3$. This implies that in the 
RD or MD evolutionary phases, 
$\rho\propto\eta^{2/(1+3w_{M})}$ is a monotonically increasing function of conformal time. 
The evolution of the Rydberg `constant' thus explains the observed cosmological redshift. 

As defined in section II, a conformal cosmic phase is a solution of 
the classical field equations characterized by all fields scaled 
by their canonical dimension. In particular, this implies that 
$|\phi|\propto\eta^{-1}$ in the comoving frame, and $\rho_{M}\propto\eta^{-4}$. 
Our generalized continuity 
equation $(\rho_{M}/\rho^{4})'+3(1+w_{M})\mathcal{F}(\rho_{M}/\rho^{4})=0$
then implies that $w_{M}=-1$, and Eqs. (15) \& (16) result in the two algebraic constraints 
$K=a_{\star}^{4}\rho_{M\star}+\lambda(a_{\star}\rho_{\star})^{2}-\theta'^{2}_{\star}$,  and 
$K=2a_{\star}^{4}\rho_{M\star}+2\lambda(a_{\star}\rho_{\star})^{2}+\theta'^{2}_{\star}$, 
where all quantities are evaluated at some pivot time $\eta_{\star}$ deep into this conformal 
cosmic phase. Here, $\theta'_{\star}$ must be constant by virtue of Eq. (18). The difference of 
these two equations results in the constraint $K=-3\theta'^{2}_{\star}$, i.e. $K$ 
must be non-positive. Since $K$ is constant, this implies that if such a conformal 
phase existed sometime in the cosmic history, the effective mass of the scalar field 
was $\sqrt{-K/3}$ and 
$\phi=\phi_{\star}(\frac{\eta_{\star}}{\eta})e^{i\sqrt{-K/3}(\eta-\eta_{\star})}$. 
Such a conformal era, that may have existed prior to its breakdown 
at some smaller $|\eta|$, is consistent only with $K\leq 0$ -- indeed the 
case considered in our cosmological scenario.

As to the frame-independence of the model at the background level, 
it is clear from Eqs. (15) or (18) that the relative evolution of all species is exactly the same as in the 
standard cosmological model, e.g. $\rho_{\gamma}/\rho_{NR}=\tilde{a}^{-1}$, 
$\rho_{\Lambda}/\rho_{NR}=\tilde{a}^{3}$. This implies that the sequence of cosmological eras, e.g. 
RD, MD, and vacuum-dominated, takes place in arbitrary frames at the same order and duration as in 
the standard cosmological model. Ultimately, any equation, e.g. Saha equation for the ionization fraction 
of hydrogen atoms, and the chemical equilibrium equations for BBN, can always be represented in 
terms of dimensionless ratios, and since the latter are unchanged in going from the 
standard cosmological model to an arbitrary field frame all 
cosmological landmarks of the former, e.g. BBN, recombination of the CMB, and recent 
accelerated expansion, are exactly mirrored in an arbitrary frame, as we now illustrate.

Consider recombination history for example. 
Since the ionization fraction of atoms $X_{e}(\eta)$ is governed by the dimensionless 
ratio of the typical wavelength of a CMB photon and the inverse Rydberg constant 
at a given time, along with the entropy per baryon, 
and since both ratios are invariant to transformation between the two frames,  
this element of cosmic recombination is unchanged in going between frames. 
The optical depth out to time $\eta_{*}$ is defined in the standard cosmological model 
as $\tau_{*}\equiv\int_{\eta_{0}}^{\eta_{*}}n_{e}X_{e}(\eta)\sigma_{T}c\cdot dt$, and since 
$n_{e}\propto a^{-3}$ and $dt\equiv ad\eta$, while $\sigma_{T}=const.$, 
the integrand is $\propto a^{-2}(\eta)$. In an arbitrary field frame
$\tau_{*}\equiv\int_{\eta_{0}}^{\eta_{*}}n_{e}(\eta)X_{e}(\eta)\sigma_{T}c\cdot d\eta$, 
the number density $n_{e}\propto a^{-3}$, but $\sigma_{T}\propto|\phi|^{-2}$ 
(due to its proportionality to the square of the electron Compton wavelength), and of course 
$dt=a d\eta$. Overall, the integrand is $\propto\tilde{a}^{-2}$, and since it 
has been shown that $\tilde{a}$ 
evolves in an arbitrary field frame exactly as does $a$ in the 
standard cosmological model, the evolution of the optical 
depth is indifferent to the frame used. A similar argument for equivalence 
with the (specific) comoving frame has been discussed in [77, 78]. Consequently, the visibility 
function $\dot{\tau}e^{-\tau}$ is frame-independent and recombination history 
is unchanged. In other words, the invariance of optical depth between the frames, rests on 
using different time coordinates in the two frames and the dilution of number 
densities in the standard cosmological model 
which is mimicked by contraction of the fundamental length scales. 
A similar equivalence argument applies to BBN history. The compatibility of our model 
with BBN is particularly important in face of claims that the FRW spacetime 
obtained from fourth-order Weyl-gravity [62] is in conflict with BBN [79]. 

This completes our discussion of the equivalence of the (background) cosmological 
models in all frames. In the next section we show that this equivalence extends also to linear 
order perturbation theory.

\subsection{Linear Perturbation Theory}
The standard cosmological model has successfully passed numerous tests and has been quite 
successful in explaining the formation and linear growth of density perturbations over the 
background spacetime, predicting the CMB acoustic peaks, polarization spectrum, and 
damping features on small scales. It also correctly describes the linear and nonlinear 
evolution phases of the LSS (on sufficiently large scales) and abundance of 
galaxy and galaxy cluster halos. 
Therefore, it would seem essential that we establish equivalence of linear 
perturbation theory between the field frames. 

In this section the equivalence of the evolution of linear scalar, vector, 
and tensor perturbations in arbitrary frames is demonstrated, i.e. 
the $a-\rho$ degeneracy found for the background evolution 
in the previous section is preserved at linear order perturbation theory. We briefly comment on 
the equivalence at higher order perturbation theory and its implication to primordial phase transitions. 
This section ends with a brief discussion of the Newtonian limit.

As in section III.C, we assume an effective matter density $\rho_{M}$ characterized by a (generally 
time-dependent) EOS $w_{M}=w_{M}(\eta)$ that encapsulates NR and relativistic baryons, DM, and radiation, 
in addition to, possibly, a potential energy density $V=\lambda\phi^{4}$. 
In the following we rewrite the perturbation equations in a manifestly $a-\rho$ symmetric fashion. Specifically,  
we show that with our conformally coupled dilatonic model $F=\rho^{2}/3$, $g_{IJ}=-2(1,\rho^{2})$, the perturbations of 
the modulus field $\delta\rho$ can be gauged 
away by redefining $\alpha\equiv\tilde{\alpha}-\delta_{\rho}$, 
$\varphi\equiv\tilde{\varphi}-\delta_{\rho}$, $\delta_{\rho_{M}}\equiv\tilde{\delta}_{\rho_{M}}+4\delta_{\rho}$,  
and $\delta_{P_{M}}\equiv\tilde{\delta}_{P_{M}}-4\delta_{\rho}$. Here, $\varphi$ and $\alpha$ 
are the Newtonian and curvature gravitational potentials, respectively, 
appearing in the perturbed FRW line element
$d\tilde{s}^{2}=\tilde{a}^{2}[(1+2\alpha)d\eta^{2}-(1+2\varphi)\gamma_{ij}dx^{i}dx^{j}]$ 
where $\gamma_{ij}\equiv diag[1/(1-Kr^{2}),r^{2},r^{2}\sin^{2}\theta]$, 
and Latin indices here run over space coordinates.
The fractional energy density and pressure perturbation in energy density units 
are $\delta_{\rho_{M}}\equiv\delta\rho_{M}/\rho_{M}$ and 
$\delta_{P_{M}}\equiv\delta P_{M}/\rho_{M}$, respectively. 
The shear $\chi\rightarrow\chi/a$, and the matter velocity $v$ remain 
unchanged. The anisotropic stress is also redefined, $\tilde{\pi}^{(s)}\equiv a^{2}\pi^{(s)}/\rho^{2}$.
The above transformations between the perturbed quantities in different field frames, 
e.g. $\alpha\equiv\tilde{\alpha}-\delta_{\rho}$, $\varphi\equiv\tilde{\varphi}-\delta_{\rho}$, etc., 
are obtained by requiring the invariance of the generalized lagrangian of massive point particles 
$\mathcal{L}_{pp}=\frac{m}{\hbar c}\int\frac{d\tilde{s}}{d\tau}d\tau$ under small field redefinitions, 
and the transformation laws for $\delta_{\rho_{M}}$ \& $\delta_{P_{M}}$ are obtained by requiring 
the invariance of the matter sector of the FRW action under linear field redefinitions. The perturbed 
modulus $\delta\rho$ of the complex dilaton field is thus not a real degree of freedom in 
{\it conformal} dilatonic gravity, much like $\delta a$ identically vanishes 
in standard FRW cosmology. This has far reaching implications, as we will see 
in subsequent sections.

Next, we write down the dynamical equations governing the evolution of 
metric and matter perturbations in a manifestly (field) frame-independent fashion, as we did 
in the previous section for the background equations. 
The linear order scalar perturbation equations are summarized in Appendix A. 
We obtain for the Arnowitt-Deser-Misner (ADM) 
energy (time-time), momentum (time-space), and propagation (space-space) components, 
Eqs. (A5), (A6) \& (A7), respectively
\begin{eqnarray}
&&(k^{2}/3-K)\tilde{\varphi}-(\tilde{\mathcal{H}}^{2}+\theta'^{2})\tilde{\alpha}
+\tilde{\mathcal{H}}(\tilde{\varphi}'-k^{2}\tilde{\chi}/3)\nonumber\\
&=&\frac{\tilde{\rho}_{M}}{2}\tilde{\delta}_{\rho_{M}}-\theta'\delta\theta'\\
&&\tilde{\mathcal{H}}\tilde{\alpha}+K\tilde{\chi}=\frac{3(1+w_{M})}{2}\tilde{\rho}_{M}u-3\theta'\delta\theta\\
&&\tilde{\chi}'+2\tilde{\mathcal{H}}\tilde{\chi}-\tilde{\alpha}-\tilde{\varphi}=3\tilde{\pi}^{(s)}/k^{2},
\end{eqnarray}
where we defined $u\equiv v/k$.
The perturbed Raychaudhuri equation (Eq. A.9) is given by
\begin{eqnarray}
&&\tilde{\mathcal{H}}\tilde{\alpha}'+(2\tilde{\mathcal{H}}'-4\theta'^{2}-k^{2}/3)\tilde{\alpha}
+k^{2}\chi'/3+\tilde{\mathcal{H}}k^{2}\chi/3\nonumber\\
&=&\tilde{\rho}_{M}(\tilde{\delta}_{\rho_{M}}+3\tilde{\delta}_{P_{M}})/2
+\tilde{\varphi}''+\tilde{\mathcal{H}}\tilde{\varphi}'-4\theta'\delta\theta'.
\end{eqnarray}
There are two scalar field equations, for $\delta\rho$ and $\delta\theta$ (Eqs. A10). 
The difference of the former equation and the perturbed trace of the Einstein tensor, 
Eq. (A11), implies that 
\begin{eqnarray}
\delta_{P_{M}}=w_{M}\delta_{\rho_{M}},
\end{eqnarray}
i.e. energy density and pressure are perturbed in a fashion that does not 
alter the EOS. We discuss 
the relation of this result to adiabatic perturbations latter in this section.
It is easy to show that the above simple relation between $\delta_{P_{M}}$ 
and $\delta_{\rho_{M}}$ is special to $\zeta=-1/6$. We emphasize that this is 
a unique property of the case $\zeta=-1/6$. In the general case 
adiabacity is lost and longitudinal field perturbations, $\delta\rho$, 
do not generally vanish.
The equation governing the scalar field phase is compactly written as
\begin{eqnarray}
\delta\theta''+2\tilde{\mathcal{H}}\delta\theta'+k^{2}\delta\theta
=\theta'(k^{2}\tilde{\chi}-3\tilde{\varphi}'+\tilde{\alpha}').
\end{eqnarray}
Finally, the continuity (A. 13) and Euler (A. 14) equations are, respectively,
\begin{eqnarray}
\tilde{\delta}'_{\rho_{M}}&=&(1+w_{M})(k^{2}\tilde{\chi}-3\tilde{\varphi}'-k^{2}u)\\
u'&+&(1-3w_{M})\tilde{\mathcal{H}}u+\frac{w'_{M}u}{1+w_{M}}\nonumber\\
&=&\tilde{\alpha}+\frac{w_{M}\tilde{\delta}_{\rho_{M}}}{1+w_{M}}
-\frac{2}{3}\left(\frac{1-3K/k^{2}}{1+w_{M}}\right)\frac{\tilde{\pi}^{(s)}}{\tilde{\rho}_{M}}.
\end{eqnarray}
Eqs. (19)-(26) summarize our linear perturbation equations; they 
are all degenerate in $a-\rho$, as is evident from the fact 
that $a$ \& $\rho$ appear only in the combinations $\tilde{a}$ 
\& $\tilde{\mathcal{H}}$. We note that $\delta\rho$ 
dropped out from the equations, and this system of eight equations (not all 
are independent) governs the evolution of $\tilde{\varphi}$, $\tilde{\alpha}$, 
$v$, $\tilde{\delta}_{\rho_{M}}$, $\tilde{\delta}_{P_{M}}$, and $\tilde{\chi}$.

Although the discussion above was limited to first order perturbations, 
there is no reason to suspect that second order modulus perturbations are not similarly 
absorbed in metric and matter perturbations. After all, the fundamental conformal action, 
Eq. (8), is obtained by requiring its invariance under 
$g_{\mu\nu}\rightarrow\Omega^{2}(x)g_{\mu\nu}$, $\phi\rightarrow\phi/\Omega$, and 
$\mathcal{L}_{M}\rightarrow\mathcal{L}_{M}/\Omega^{4}$, for {\it arbitrary} $\Omega(x)$.
In other words, $\rho$ and its perturbations only represent our freedom to locally gauge 
the metric-, Dirac-, and gauge-fields, in conformal dilatonic gravity (the case $\zeta=-1/6$).
This observation is significant as the theory of primordial phase transitions hinges on 
the existence of the non-vanishing of (at least) second-order field perturbations; their rms 
represent thermal fluctuations that depend on the temperature, and the effective potential 
becomes temperature-dependent. This is essential for the proper working of electroweak and 
QCD phase transitions at $O(100)$ GeV and $O(200)$ MeV, respectively, as well as possibly other 
phase transitions, at the pre-BBN era. No clear traces of these transitions have 
been so far identified in observational data. These could potentially include topological 
defects such as magnetic monopoles, cosmic strings, and domain walls -- none has 
been observed. Primordial phase transitions are an essential theoretical component of our 
understanding within the standard cosmological model framework 
of baryon and lepton asymmetries, e.g. [80, 81]. 
Thus, the underlying conformal invariance of the theory considered here by its very essence 
rules out cosmological phase transitions due to thermal fluctuations of scalar fields, unless 
one or a few of them are not conformally coupled to gravity. Our working assumption is that 
conformal invariance is an {\it exact symmetry} of the fundamental interactions.

Vector and tensor perturbations are described by Eqs. (53)-(55) and (58) of [38] respectively, 
and the equivalence between the models is straightforward to show in this case, provided 
that $a\rightarrow\tilde{a}$ and $\rho_{M}\rightarrow\rho_{M}/\rho^{4}$ 
and $\pi^{(v),(t)}\rightarrow\pi^{(v),(t)}/\rho^{4}$, 
where $\pi^{(v)}$ and $\pi^{(t)}$ are the stresses associated with 
vector and tensor mode perturbations, respectively. 

The full kinetic theory, pertaining to the theory described by Eq. (1), 
involving collisional photons and collisionless 
neutrinos is given in [38], where the corresponding perturbed energy-momentum tensors are 
given in terms of integrals over the respective distribution functions $f$. The latter are 
obtained in the standard way by integration of the Boltzmann equations over perturbed 
FRW backgrounds. In the case of neutrinos $f_{\nu}=f_{\nu}(x^{\mu},p^{\mu},M_{\nu})$, 
where $x^{\mu}$, $p^{\mu}$ and $M_{\nu}$ are spacetime coordinates, canonical momenta, 
and neutrino mass, respectively. 
In our case, however, $M_{\nu}$ is a field; more 
explicitly, in our comoving frame example the comoving neutrino energy at momentum $p$ 
would be $E_{\nu}(q)=\sqrt{q^{2}+(aM_{\nu}c)^{2}}$, where $p=q/a$, and in our notation 
$E_{\nu}(q)=\sqrt{q^{2}+(\tilde{a}\lambda_{\nu})^{2}}$. Here, $\lambda_{\nu}$ is the Planck 
length measured in neutrino Compton wavelength units.

The Newtonian limit of conformal dilatonic gravity 
is obtained from Eqs. (19)-(26) by setting $\tilde{\mathcal{H}}$, $K$, $w_{M}$, 
and $\theta'$ to 0. In particular, Eqs. (19) \& (26) are the Poisson, continuity, and 
Euler equations, respectively. For comparison, obtaining the Newtonian 
limit of fourth order Weyl gravity is significantly less trivial [82-84].

By virtue of the frame equivalence discussed in this and the previous section 
the null geodesics on the perturbed FRW geometry 
followed by photons are unchanged in transforming between frames. 
This would imply, in particular, 
that once CMB perturbations are shaped up at the last scattering surface their 
propagation towards an observer is not altered in transforming between field frames. 
Similarly, timelike geodesics are unaffected and matter infall and 
structure formation histories are unchanged. 
Indeed, as mentioned earlier, world lines of 
a massive point particle are obtained from 
$\mathcal{L}\propto m\int \frac{ds}{d\mu}d\mu$, 
i.e. rescaling of $g_{\mu\nu}$ is offset by rescaling of $m$, 
or rather $mc/\hbar=l_{C}^{-1}$. In the following discussion it 
is understood that the gauge $\delta\rho=0$ is chosen and 
the tilde symbols are dropped.

Eq. (23) implies that matter density perturbations 
and pressure perturbations must be concerted in a fashion that 
does not locally vary the EOS of matter, i.e. the speed of sound, at all times. 
In other words, entropy is not perturbed, and this holds true for any combination of 
species in our effective single-fluid description of matter with $w_{M}(\eta)$.
The entropy perturbation $\Gamma_{i}$ associated with the i'th species is defined 
via $w_{i}\Gamma_{i}\equiv\delta_{p_{M,i}}-c_{i}^{2}\delta_{\rho_{M,i}}$ 
where $c_{i}^{2}\equiv P'_{M,i}/\rho'_{M,i}$ defines the sound speed of the i'th fluid. 
In other words, for this relation to hold at all times, with temporally-varying 
weights of the various species (e.g. baryons, DM, radiation, DE), 
this implies that matter perturbations must be adiabatic. 

To prove the above claim,  
we consider the total entropy of two species, the intrinsic entropy of each vanishes initially. 
It is straightforward to show that the total entropy of a two-component fluid is (see, e.g. [85])
$w(1+w)\Gamma_{tot}=R(1-R)(c_{1}^{2}-c_{2}^{2})\left[(1+w_{2})\delta_{1}-(1+w_{1})\delta_{2}\right]$ 
where $R\equiv\rho_{2}/\rho$ \& $1-R\equiv\rho_{1}/\rho$, and $c_{i}$ \& $\delta_{i}$ 
are the sound speed and fractional energy density of the i'th species, and therefore for Eq. (23) to 
hold at all times, assuming that neither of the species is subdominant and the sound speeds are 
not necessarily the same, i.e. $c_{1}\neq c_{2}$, 
then the adiabacity condition $\delta_{1}/(1+w_{1})=\delta_{2}/(1+w_{2})$ is obtained. 
This implies that for a multi-component fluid Eq. (23) is only consistent with 
adiabatic perturbations.

By virtue of the discussion below Eq. (18), and assuming that the $\rho_{M,i}/|\phi|^{4}$ are 
indeed conserved for each species $i$ separately, then 
the average number density of the i'th species characterized by an EOS $w_{i}$ is
$n_{M,i}=\left(\rho_{i}/(\hbar c)\right)^{\frac{1}{1+w_{i}}}\rho^{-\frac{1-3w_{i}}{1+w_{i}}}$.
Recalling that $\delta\rho=0$, and perturbing this relation we obtain the 
fractional perturbation of the number density of the i'th species
$\frac{\delta n_{i}}{n_{i}}=\frac{\delta_{\rho_{i}}}{1+w_{i}}$.
The first analysis [33] to show that the conformal phase of the theory described by 
Eq. (8), albeit with no matter present, with $\theta'=0$, flat spacetime, 
and no metric perturbations implies 
not only that $\delta_{\rho}$ is generally non-vanishing, but that it 
is also characterized by a red power spectrum [33]. 
An additional assumption had to be made in order to suppress its effect on relevant 
cosmological scales.
In the previous section it was shown that with the inclusion of 
metric perturbations, the fluctuations $\delta_{\rho}$ are spurious anyway. 
Other issues with the treatment of [33] are inherent non-gaussianity 
and statistical anisotropy in scalar perturbations. Both are due to higher order 
couplings of $\delta\rho$ to $\delta\theta$ [33, 86, 87] which vanish in our treatment 
due to gauging $\delta\rho$ away, as is described above. In other words, 
and from the viewpoint taken in present framework, the issues raised in [33] are due to gauge 
artifacts. 

\subsection{Very Early Universe Scenario}
As a consequence of the results of section III.D, real field inflation cannot be responsible 
for the seed metric perturbations in conformal dilatonic cosmology 
since modulus perturbations identically vanish. This is contrary to the presumed scenario in 
the standard cosmological model, namely that modulus perturbations are quantum mechanically 
excited and ultimately decay into ordinary matter ($\delta\rho_{M}\approx\frac{dV}{d\Phi}\delta\Phi$). 
Thus, inflation can explain away the classical problems that it was designed to address, but it cannot 
explain metric perturbations in conformal dilatonic theory.

According to the alternative scenario proposed here a single complex scalar field 
does not only resolve the flatness and horizon puzzles, but also accounts for scale-free density 
perturbations. 
The relic topological problem (the `magnetic monopole problem') does not arise (in 
a conformal dilatonic theory of the fundamental interactions) in the first place because 
primordial phase transitions do not occur. 
This scenario assumes $\theta'\neq 0$, i.e. the energy 
density associated with the evolving dilaton phase, effectively a stiff matter, 
is dynamically relevant at very small modulii values, which would correspond 
in the standard cosmological model to the would-be (or near the) initial singularity. 

The proposed scenario is non-singular and the cosmic history comprises 
contracting and expanding evolution phases (as would be described in the EF). 
The former is dominated by a conformal era (akin to a `deflationary' epoch) 
characterized by $w=-1$, followed by (possibly) a curvature-dominated (CD), MD, RD, and an 
effectively stiff matter dominated eras, a bounce, and an expansion with these 
various pre-bounce eras occurring in reverse order. The model is adiabatic; 
matter is not created nor is it destroyed throughout cosmic history.

Assuming a nonvanishing (negative) curvature, that ordinary matter was relativistic in 
the very early universe, and considering the quartic potential term as well, 
Eq. (15) implies that 
\begin{eqnarray}
\mathcal{H}^{2}+K=\lambda\tilde{a}^{2}+\rho_{r,*}/\tilde{a}^2+\rho_{\theta,*}/\tilde{a}^{4},
\end{eqnarray}
where $a_{*}\equiv 1$, $\rho_{i,*}$ is the energy density associated with the i'th species 
at $\eta_{*}$, $\rho_{\theta}$ is the effective {\it negative} energy density associated with the 
dynamical phase $\theta$, and $\lambda>0$.
The full analytic integration of this equation is not very illuminating, and therefore we 
treat two interesting limits separately. This will suffice for our purposes.  
The first limit is obtained by neglecting the $\lambda\tilde{a}^{2}$ term which 
is negligible compared to the other terms in the limit of very small $\tilde{a}$. 
In this case, the Friedmann-like equation integrates to
\begin{eqnarray}
\tilde{a}^{2}&\approx&\left(\frac{\rho_{\theta,*}}{K}+
\left(\frac{\rho_{r,*}}{2K}\right)^{2}\right)^{1/2}\cosh\left(2\sqrt{-K}(\eta-\eta_{*})\right)\nonumber\\
&+&\frac{\rho_{r,*}}{2K},
\end{eqnarray}
where $\tilde{a}$ attains its minimum at $\eta=\eta_{*}$.
If, in addition, the curvature identically vanishes then Eq. (27) 
integrates to $\tilde{a}^{2}=\rho_{r,*}(\eta-\eta_{*})^{2}-\rho_{\theta,*}/\rho_{r,*}$.
The latter could be readily integrated to give the cosmic time around the bounce, 
$t=\int a(\eta)d\eta$, and is easily verified to be non-singular as well. 
In other words, the effective time coordinates of both massless and massive particles 
can be extended through the bounce to $-\infty$. Thus, associating the Planck mass 
with the modulus of a complex scalar field results in a possible non-singular 
bouncing cosmological model due to the non-trivial interplay between the 
evolutions of the dilaton phase encapsulated by $\rho_{\theta}$ 
(which acts as an effective stiff fluid with negative energy density), 
and the positive energy densities carried by (relativistic) matter 
and (negative) curvature. The causal horizon can thus be 
made arbitrarily large, thereby avoiding any horizon problem.
To guarantee thermalization of the CMB the minimal $\tilde{a}$ at 
the bounce $\sqrt{|\rho_{\theta,*}|/\rho_{r,*}}$, and the 
expansion rate around it, must allow sufficient time for 
efficient (particle-number non-conserving) 
double-Compton and bremsstrahlung interactions. 
This can always be achieved for negligibly small 
$|\rho_{\theta,*}|$ where the background evolution asymptotically 
converges to the standard cosmological model. The more interesting 
case of larger $|\rho_{\theta,*}|$ is constrained by BBN, in addition 
to the required CMB thermalization. 

At the other extreme, that of a sufficiently large $\tilde{a}$, neglecting the stiff 
matter and radiation contributions, the Friedmann equation integrates to   
\begin{eqnarray}
\tilde{a}=\sqrt{\frac{-K}{\lambda}}\frac{1}{\sinh\left(\sqrt{-K}(\eta'-\eta)\right)},
\end{eqnarray}
where $\eta'$ is an integration constant and the domain of validity of Eqs. (28) \& (29) 
do not overlap. Positivity of $\tilde{a}$ implies that $\eta<\eta'$.
This solution has the correct behavior at the $\sqrt{-K}(\eta'-\eta)\gg 1$ 
and $\sqrt{-K}(\eta'-\eta)\ll 1$ limits, which correspond to the CD 
and conformal epochs, respectively. 

In the latter case (an evolution phase dominated by the quartic potential) 
$\tilde{a}=\left(\sqrt{\lambda}(\eta'-\eta)\right)^{-1}$, and $\tilde{a}$ scales according 
to its canonical dimension $length^{-1}$, i.e. $\propto\eta^{-1}$, not $\propto t^{-1}$. 
This again highlights the privileged 
role played by conformal time as compared to cosmic time, in contrast to the standard 
cosmological model where conformal time is only used for computational convenience, 
or as the natural time coordinate parameterizing null geodesics.
We see that during the conformal era $\tilde a$ behaves in the `deflationary' epoch 
exactly as does $a$ during the inflationary era. From this perspective, the `slow-roll' condition 
(i.e. the near constancy of the inflaton field during the inflationary era of the standard 
cosmological model) could then be understood merely as the EF 
version of our frame-independent solution, $\tilde{a}=\left(\sqrt{\lambda}(\eta'-\eta)\right)^{-1}$, 
in which scalar fields of units $mass^{-1}$ are fixed, much like particle masses and $G$ are.

The `flatness problem' arises in the hot big bang model due to the monotonic 
expansion of space and the consequent faster dilution of the energy density 
of matter (either relativistic or NR) than the effective energy density dilution 
associated with curvature. It is thus hard 
to envisage how could space be nearly flat (as is indeed inferred from observations) 
if not for an enormous fine-tuning at the very early universe, or an 
early violent inflationary era. 
In the proposed bounce scenario the matter content of the universe has always 
existed, and in particular the present ratio of matter- to curvature-energy densities 
has been exactly the same when $\tilde{a}$ in the pre-bounce was equal to its 
present value. However, in the pre-bounce phase matter domination over curvature 
is actually an {\it attractor point} as the contracting universe has no beginning. 
In other words, had the universe been CD at present (as is naively expected in the standard 
{\it expanding} cosmological model with no inflation), i.e. at $\tilde{a}_{0}$, it must have been CD 
at $\tilde{a}_{0}$ at the pre-bounce phase, but since $\rho_{M}$ grows faster than $\rho_{K}$ 
in this phase then a CD domination would amount to an extremely 
fine-tuned $\rho_{M}/\rho_{K}\rightarrow 0$ at $\eta\rightarrow-\infty$.
As is well-known, entropy produced in the pre-bounce era is processed 
at the bounce to thermal radiation, implying in effect that $\rho_{r}$ might somewhat 
change between pre- and post-bounce but the expectation that the contracting and 
expanding phases nearly mirror one another is not significantly changed -- not at any rate 
that might change the conclusion regarding the (un-) naturalness of CD era at the present.

As we show below, primordial scalar perturbations are seeded during 
the conformal era at the pre-bounce phase. During that phase the field 
kinetic and potential terms differ only by sign, and contrary to the case of inflation in 
the standard cosmological model, the dilaton here does not slow roll. Therefore, 
fine-tuning of the inflationary potential against quantum corrections that could potentially 
spoil the required flatness of the potential (the `$\eta$-problem') 
is not required in our scenario.
On the contrary, conjecturing conformal symmetry implies that the only allowed 
(purely) self-interaction of scalar fields is described by quartic 
potential, which indeed results in this conformal evolution phase. 
In fact, it should be clear from section III.C that the vacuum-like energy 
required for inflation, given in Planck units, is exactly the dimensionless coupling 
constant in the $\lambda|\phi|^{4}$ potential of Eq. (8). Perhaps ironically, 
the `slow-roll' condition may thus be interpreted as a hint for the (unbroken) 
conformal nature of gravity in the presence of scalar fields {\it only}. 

Since $\delta\rho$, the perturbed modulus field, 
is non-dynamical due to the local scaling invariance it represents, 
and is therefore set to vanish, (simply a gauge choice as discussed 
in section III.D), then only transversal perturbations of our scalar 
field $\phi$ could potentially source density 
perturbations. Indeed, the latter, as opposed to $\delta\rho$, 
is only {\it minimally-coupled} -- rather than conformally-coupled -- 
to gravity. Working in the shear-free gauge $\chi=0$, 
and neglecting stress anisotropy, Eq. (21) implies $\varphi=-\alpha$. 
We obtain from section III.D the following set of 
linear perturbation equations in an arbitrary frame (the ADM energy constraint, 
the momentum constraint, the perturbed Raychaudhuri equation, the perturbed equations 
for the modulus and phase of the scalar field, the perturbed continuity, and Euler equations, 
respectively) 
\begin{eqnarray}
&&\tilde{\mathcal{H}}\varphi'+\left(\frac{k^{2}}{3}-2K+\tilde{a}^{2}\tilde{\rho}_{M}\right)\varphi\nonumber\\
&=&\frac{\tilde{\rho}_{M}}{2}\delta_{\tilde{\rho}_{M}}-\theta'\delta\theta'\\
&&\varphi'+\tilde{\mathcal{H}}\varphi=-\frac{3(1+w_{M})\tilde{\rho}_{M}u}{2}+3\theta'\delta\theta\\
&&\varphi''+2\tilde{\mathcal{H}}\varphi'-\left[\frac{k^{2}}{3}
+(1+3w_{M})\tilde{a}^{2}\tilde{\rho}_{M}\right]\varphi\nonumber\\
&=&-\frac{(1+3w_{M})\tilde{\rho}_{M}\delta_{\tilde{\rho}_{M}}}{2}+4\theta'\delta\theta'\\
&&\varphi''+4\tilde{\mathcal{H}}\varphi'+\left(\frac{k^{2}}{3}-4K
+4\tilde{a}^{2}\tilde{\rho}_{M}\right)\varphi\nonumber\\
&=&2\theta'\delta\theta'\\
&&\delta\theta''+2\tilde{\mathcal{H}}\delta\theta'+k^{2}\delta\theta=-4\theta'\varphi'\\
&&\delta'_{\tilde{\rho}_{M}}=-(1+w_{M})(3\varphi'+k^{2}u)\\
&&(1+w_{M})u'+(1+w_{M})(1-3w_{M})\tilde{\mathcal{H}}u\nonumber\\
&+&w'_{M}u=(1+w_{M})\alpha+w_{M}\delta_{\tilde{\rho}_{M}}.
\end{eqnarray}
In the special case of $w_{M}=-1$, it follows that 
$\delta_{\tilde{\rho}_{M}}=0=u$. 
When $\theta'=0$ the transversal perturbation mode is decoupled from the other perturbations 
and is fully described by Eq. (34). The appropriate Bunch-Davis vacuum imposed in the $k\eta\gg 1$ 
limit $\delta\theta\rightarrow \frac{e^{-ik\eta}}{\sqrt{2k}}$ 
implies that $\delta\theta\propto H^{(1)}_{3/2}(k\eta)$, where $H^{(1)}$ is 
a Hankel function of the first kind, which corresponds to a scale-free power spectrum 
$P_{\delta\theta}(k)=constant$ in the large scale limit $k\eta\ll 1$, exactly as in [33].

The above analysis is now generalized by considering the direct coupling 
between the scalar field fluctuations and metric perturbations, along with allowing 
a spatially curved background. 
Turning metric perturbations $\varphi$ on, and in addition allowing for $\theta'\neq 0$,
Eqs. (30)-(36) are all consistent with a single equation
\begin{eqnarray}
\varphi''+6\tilde{\mathcal{H}}\varphi'+(q^{2}-6\lambda\tilde{a}^{2})\varphi=0 ,
\end{eqnarray}
where $q^{2}\equiv k^{2}-8K$. Note that $\theta'$ vanishes from Eq. (37) although 
it played a crucial role in coupling metric perturbations to $\delta\theta$ 
(as is seen from Eq. 34). We are interested in a solution in the conformal 
phase where $\tilde{a}=1/(\sqrt{\lambda}\eta)$. 
Neglecting spatial curvature, the general solution of Eq. (37) is 
$\varphi=c_{1}\eta^{\frac{7}{2}}J_{5/2}(k\eta)+c_{2}\eta^{\frac{7}{2}}Y_{5/2}(k\eta)$.
Eq. (31) implies that $\delta\theta=(\varphi'-\varphi/\eta)/(3\theta')$. 
Again, requiring the appropriate Bunch-Davis vacuum for $\delta\theta$ at $k\eta\gg 1$ 
determines the coefficients $c_{1}$ and $c_{2}$ and results 
in $\varphi\propto k^{-1}\eta^{5/2}H_{5/2}^{(1)}(k\eta)$, which in the $k\eta\ll 1$ 
limit corresponds to $k^{3}P_{\varphi}(k)=k^{-4}$. This translates 
to $k^{3}P_{\delta\rho_{M}}(k)\equiv k^{3}P_{M}(k)=constant$, 
a flat matter power spectrum, by virtue of the Poisson equation. Although inflation 
provides a mechanism for generating scalar and tensor perturbations which are characterized 
by a nearly-flat power spectrum, it is not a prediction of the inflationary scenario; it 
has been known for nearly a decade before the advent of inflation that at least the density 
perturbations are described by a nearly flat spectrum [88, 89]. Other early universe 
scenarios, e.g. the varying speed of light cosmology [90, 91], the ekpyrotic [92] 
and new ekpyrotic [93] scenarios, the cyclic universe [94, 95], string gas cosmology [96], 
Anamorphic cosmology [97], and pseudo-conformal universe [33, 98], are 
capable of explaining the observed flat spectrum as well.

It should be stressed that $\delta\theta$ appears only in the kinetic term in 
the perturbed lagrangian; it does not appear in either the potential or the non-minimal 
coupling terms due to the $U(1)$ symmetry. Scalar perturbations 
are therefore generated by the kinetic term, unlike in standard inflation where primordial 
scalar perturbations are produced by fluctuations in the potential term.

As was argued in [33], the existence of both the 
$k(\eta_{i}-\eta')\gg 1$ and $k(\eta_{f}-\eta')\ll 1$ limits in Eq. (37) 
(where $\eta_{i}$ and $\eta_{f}$ are conformal time at the onset and 
at the end of the conformal era, respectively) requires 
that $\eta_{f}-\eta_{i}$ should be much larger than a few Gpc, the 
comoving Hubble scale. This condition is trivially satisfied during the 
(infinitely old) contracting evolution phase in the non-singular model.
We reiterate in this context that the non-singular nature of the cosmological 
model is due to the negative contribution of the dynamical dilaton 
phase to the total energy density, a contribution neglected in [33]. 
Once again, we see that invoking complex 
rather than a real field is vital in conformal dilatonic gravity; not only that the 
quantum fluctuations of the phase results in a nearly flat power spectrum of scalar metric 
perturbations, the dynamics of the phase itself guarantees that the conformal era 
lasts for sufficiently long time to explain the spectrum on cosmological scales, and at 
the same time avoids the initial singularity. We stress that the discussion below Eq. (9) 
addressed the general property of curvature scalars that transform inhomogeneously 
under conformal transformations and its potential use in removing curvature singularities 
at the cost of introducing singularities in the scalar fields. This should be distinguished 
from the above discussion where a specific model is considered that is entirely non-singular, 
in both scalar fields and curvature scalar.
The (vacuum-dominated) conformal phase must cease at some point and ultimately give way 
for radiation-dominated (RD) followed by matter-dominated (MD) epochs 
to maintain consistency with light element abundance and the LSS. 

Since the underlying mechanism outlined here 
for the generation of initial perturbation involves the quantum vacuum state, it is clear that the 
perturbations are gaussian. Now, although $\lambda$ does not appear in the final expressions,  
it is clear that turning this dimensionless coupling off we would not have obtained 
a conformal evolution phase to begin with. 
The effective mass of the Newtonian potential field (Eq. 37) is 
$m_{\varphi}^{2}=-8K+6/\eta^{2}$, which vanishes as $\eta\rightarrow -\infty$ 
(given the observation that the spatial curvature radius is super-Hubble scale). 
Therefore, the field $\varphi$ is at best very weakly self-coupled 
and one expects that the gaussianity of scalar metric perturbations is not 
spoiled by strong self-interaction. 

The power spectrum inferred from cosmological observations is not strictly flat. 
This slightly red-tilted spectrum might be obtained by 
mildly breaking the underlying conformal symmetry 
of the model. It has been proposed in [99] that this 
can be achieved by considering the potential 
$V\propto\rho^{4+\alpha}$ where $\alpha$ is a small tilt. 
Our approach is somewhat similar, at least in the sense that to generate a red-tilted power 
spectrum we break away from the $w_{M}=-1$ case. By virtue of Eq. (7), in a theory containing 
not only scalar fields, $w_{M}\neq -1$ {\it does not} imply breakdown of conformal symmetry 
at the classical action level. The analysis described below selects the 
tilt sign by a sound physical criterion (rather than making an 
{\it ad hoc} choice of, e.g. the sign of $\alpha$), and naturally avoids the ghost problem discussed 
in section III for scalar fields with negative kinetic energy. To achieve this we assume an arbitrary 
species with a nearly constant EOS $w_{M}\approx -1$. 
From the above discussion it is clear that the 
resulting power spectrum of density perturbations is nearly flat; this is derived below. 
As is seen from Eq. (7), in the absence of non-scalar fields the only possible EOS is 
$-1$. Therefore, the near flatness of power spectrum 
(rather than it being strictly flat) is tightly connected to the presence 
of other fields, e.g. Dirac fields, in the model. In comparison, the tilts of scalar/tensor 
perturbations spectra in standard cosmology inflation depend only on the shape (convexity) of the 
inflaton potential and by no means require the presence of other fields. 
The latter are supposedly decay products of the inflaton field itself during reheating.

For a $w_{M}=constant$, we found earlier 
that $\tilde{\rho}_{M}=\tilde{\rho}_{M,0}\tilde{a}^{-3(1+w_{M})}$  
which implies -- by virtue of the generalized Friedmann equation -- that 
$\tilde{\mathcal{H}}=\frac{2}{(1+3w_{M})\eta}$. Eq. (37) then becomes
\begin{eqnarray}
\varphi''-\frac{6}{\mathcal{C}}\frac{\varphi'}{\eta}
+\left(k^{2}-\frac{6}{\mathcal{C}^{2}\eta^{2}}\right)\varphi=0, 
\end{eqnarray}
where $\mathcal{C}\equiv 1-3\Delta w_{M}/2$, and $\Delta w_{M}\equiv 1+w_{M}$ 
measures the departure from $\tilde{\rho}_{M}\propto|\phi|^{4}$. 
The general solution of Eq. (38) is (considering for simplicity 
$\mathcal{C}=constant$ at the vicinity of $\mathcal{C}=1$)
\begin{eqnarray}
\varphi=c_{1}\eta^{\frac{6+\mathcal{C}}{2\mathcal{C}}}J_{\beta}(k\eta)
+c_{2}\eta^{\frac{6+\mathcal{C}}{2\mathcal{C}}}Y_{\beta}(k\eta),
\end{eqnarray}
where $\beta\equiv(12+12\mathcal{C}+\mathcal{C}^{2})^{1/2}/(2\mathcal{C})$. The term 
$-3(1+w_{M})\tilde{\rho}_{M}u/2$ in Eq. (36) is second order near $w_{M}=-1$, and 
can be neglected. Under this approximation we 
obtain, as in the $w_{M}=-1$ case, 
that $\delta\theta=(\varphi'+\tilde{\mathcal{H}}\varphi)/(3\theta')$. 
Therefore 
\begin{eqnarray}
\delta\theta(k|\eta|\gg 1)\approx\sqrt{\frac{2}{\pi k}}
\left(c_{-}d_{+}e^{ik\eta}+c_{+}d_{-}e^{-ik\eta}\right),
\end{eqnarray}
where $c_{\pm}\equiv(c_{1}\pm ic_{2})/2$ and 
$d_{\pm}\equiv \left(\frac{4+\mathcal{C}}{2\mathcal{C}}\right)
\eta^{(1-\mathcal{C})/\mathcal{C}}\pm ik\eta^{1/\mathcal{C}}$. 
By imposing a Bunch-Davis vacuum, $c_{-}=0$, we obtain that $c_{+}\propto k^{-1}$, and 
requiring that the perturbation $\delta\theta$ never diverges at $\eta\rightarrow 0$ 
(i.e. near the bounce, in the pre-bounce phase), 
we obtain the condition $0<\mathcal{C}<1$, i.e. $w_{M}\gtrsim -1$. 
In other words, if $\rho_{M}$ is predominantly 
contributed by a scalar field, its EOS must be $w_{M}\gtrsim -1$ for its perturbations to 
generate a nearly scale-free spectrum of scalar perturbations that survive cosmic evolution. 
Rather remarkably, regularity of linear perturbations at the bounce (to avoid conflict with 
the underlying assumption of a homogeneous universe during the early post-bounce phase), 
by itself, selects the quantizeable scalar fields of our model 
(i.e. those scalar fields comprising $\tilde{\rho}_{M}$, the dominant contribution 
to the energy density at the time of generation of density perturbations) to be those characterized 
by $w_{M}\gtrsim -1$, i.e. the effective scalar field description of $\tilde{\rho}_{M}$ must be 
of the canonical form with a non-negative kinetic term, e.g. [40, 100]. 
This has no immediate consequences for our derivation of the density perturbations 
as we considered classical $\tilde{\rho}_{M}$, and consequently the classical $\tilde{\mathcal{H}}$ that 
it induces, much like $\mathcal{H}$ is considered a classical field in standard cosmology. Yet, it is 
of some conceptual significance that self-consistency of the model selects the properties of matter 
to be compatible with the quantization procedure.
The dilaton field of our model, i.e. the Planck field, 
or the cosmic scale factor $a$, has negative kinetic 
energy, and does not leave any imprint from its quantized perturbations -- if any at all -- 
in the early universe. 

Next, we show that the spectrum is necessarily red-tilted. On large scales the power spectrum 
associated with the gravitational potential of Eq. (39) is $k^{3}P_{\varphi}\propto k^{1-2\beta}$, and 
in the MD era when $\varphi$ is time-independent this is related via the Poisson equation 
to the matter power spectrum, $k^{3}P_{M}\propto k^{5-2\beta}$. In the pure potential case ($w_{M}=-1$) 
$\beta=5/2$ the spectrum is back to the Harrison-Zeldovich form. 
A blue power spectrum is obtained for any $\mathcal{C}>1$ or $\mathcal{C}<-1/2$, 
and a red spectrum is obtained in the case $-1/2<\mathcal{C}<1$. 
Combining this with the condition of regularity at around the bounce, 
$0<\mathcal{C}<1$, we obtain the result 
that the scalar perturbations are generated by 
almost  purely potential 
field (effectively $w_{M}\gtrsim -1$). We see that the regularity condition by itself not only 
selects those scalar fields with canonical kinetic energy, 
it also guarantees that the power spectrum is red-tilted. For best-fit parameters of the scalar index,  
$n_{s}\approx 0.97$, e.g. [101], we obtain $w_{M}\approx -0.99$. 
In comparison, the inflationary scenario generally 
predicts a red-tilted spectrum but in the latter case the 
tilt sign is determined by the generic shape of the 
inflaton potential, i.e. the positivity of its first and second derivatives.
Since the pseudo-conformal deflationary phase of $\tilde{a}$ is nearly 
a mirror picture around 
$\tilde{a}_{min}$, i.e. the scale $\tilde{a}$ at the bounce, of the present 
pseudo-conformal (`accelerated') expansion we thus find a relation between 
$n_{s}$ and $w_{DE}$ -- a relation that does not {\it a priori} exist 
in the standard cosmological model. In other words, this relation is obtained thanks to 
the fact that in the scenario discussed here the present dominance 
of `DE' is a mirror picture around the bounce of the pre-bounce deflationary 
phase. 

The issue of negative kinetic energy associated with the 
scalar fields is characteristic of either GR or 
other theories of gravitation `dressed' with conformal symmetry, e.g. [51, 102]. This is usually interpreted 
as an indication for instability of the theory, e.g. [12, 46, 49, 102-104], and was at the heart of an 
intense 
debate regarding the physical equivalence of theories in the `EF' and `JF' where the latter was considered 
by many as `unphysical', in spite of the fact of being mathematically 
equivalent to the former, e.g. [50, 105], 
while others maintained that at least classically the theory with negative kinetic energy of the scalar field 
is stable [105-109]. Some even speculated that in a semi-classical context, when $\phi$ is 
quantized and the spacetime metric is treated as a classical field, stability may still be maintained, 
e.g. [109]. The above analysis implies that in an evolving universe, growing scalar modes from 
quantum seed perturbations associated with the fluctuating phases of non-positive kinetic terms 
must be driven by matter whose effective scalar field description is canonical, i.e. 
its kinetic term is non-negative. In contrast, scalar fields with 
non-positive kinetic terms do not result in growing scalar perturbation modes but should 
rather be treated as classical fields (at least in the cosmological context), much like 
the cosmic scale factor is considered classical. 

In the pre-bounce pseudo-conformal phase relativistic and NR matter 
contribution must be negligible by definition. Any slow-rolling scalar field 
with non-negative kinetic and potential energies could result 
in $w_{M}\gtrsim -1$ 
in the pre-bounce phase. In the far pre-bounce 
phase slow-rolling scalar fields might bear some resemblance to standard 
cosmology quintessence. For comparison, in standard cosmology inflation 
the slow-roll condition is applied at energy scales as high as the GUT scale 
where quantum corrections could potentially spoil the required flatness of the 
inflaton potential.

A generic prediction of the standard cosmological model is that PGW 
are generated during inflation. These are expected to 
leave their imprint in the CMB in the form of a typical B-mode polarization signature 
on angular degree scales on the sky. This implies that 
if B-mode polarization is ultimately observed in the CMB, the mechanism discussed here 
for the generation of density perturbations needs to be supplemented by a mechanism 
for PGW production. In contrast, if inflation did indeed occur, B-mode detection 
might still be impossible if inflation took place at a sufficiently late epoch, i.e. 
at characteristic energy scales $\lesssim 2\times 10^{15} GeV$ [110]. 
Even if inflation took place at higher energies, 
polarized galactic dust is 
expected to hinder such measurements. Thus, non-detection of B-mode polarization 
would only imply for the standard cosmological model that inflation transpired 
relatively late compared to the Planck scale, while a detection 
of a statistically significant primordial B-mode signal would likely undermine 
the mechanism proposed here (apparently the only natural 
mechanism to achieve that goal in the proposed framework) for generation of perturbations.

Since the metric perturbations at recombination are known to be of $O(10^{-5})$ and according 
to the proposed scenario they have been seeded during the pseudo-conformal era at the pre-bounce 
epoch it is legitimate to ask whether the original linear perturbations survive the bounce rather 
than develop anisotropic instabilities that could potentially undermine the assumed homogeneity 
and isotropy at the (what is normally considered) very early universe, i.e. the universe shortly 
after the bounce. Choosing the time coordinate such that at the bounce it vanishes, and 
considering the case that near the bounce the cosmic energy budget is dominated by relativistic 
degrees of freedom, $\rho_{r}$, and the effective stiff matter $\rho_{\theta}$, Eq. (27) then 
integrates to $\tilde{a}^{2}=\rho_{r,*}\eta^{2}-\rho_{\theta,*}/\rho_{r,*}$ (where we recall that 
$\rho_{\theta,*}<0$) and $\rho_{i,*}$ is the energy density associated with the i'th species at 
the bounce. Plugging this with $w_{M}=1/3$ and $K=0$ in Eqs. (30)-(36) and focusing on 
superhorizon scales, $k\eta\ll 1$, we obtain for the scalar metric perturbations
$(\rho_{r,*}\eta^{2}-\rho_{\theta,*}/\rho_{r,*})\alpha''+6\rho_{r,*}\eta\alpha'
+6\rho_{r,*}\alpha=0$.
A solution compatible with the requirement that the metric and its perturbations are real is
$\alpha=-\varphi=c_{\alpha}\eta\left(-\rho_{\theta,*}/\rho_{r,*}
+\rho_{r,*}\eta^{2}\right)^{-2}$, 
where $c_{\alpha}$ is an integration constant. 
Therefore, at the bounce $\alpha_{*}=0$ and is finite in its proximity 
(and this is allowed only thanks to $\rho_{\theta,*}\neq 0$). 
Its maximal value 
is obtained at $\eta_{max}^{2}=|\rho_{\theta,*}|/(3\rho_{r,*}^{2})$. 
Similarly, the linear perturbation equations result in 
$\delta\theta=-\sqrt{3}(\eta_{max}/\eta)\alpha$ which is also finite and obtains its maximum 
at the bounce. By choosing $c_{\alpha}$ to be sufficiently small both $\alpha$ 
and $\delta\theta$ can be kept small at the bounce and its vicinity. 
Repeating this analysis for $\xi\equiv k\eta\gg 1$, we obtain
$3\xi^{2}\ddddot{\alpha}+30\xi\dddot{\alpha}
+4\xi^{2}\ddot{\alpha}+18\xi\dot{\alpha}+\xi^{2}\alpha=0$, 
where $\dot f\equiv\frac{df}{d\xi}$. 
Imposing the condition that $\alpha$ is real results in 
$\alpha=c_{1}\cos(\omega\xi)/\xi^{2}+c_{2}\sin(\omega\xi)/\xi^{2}$ where 
$c_{1}$ and $c_{2}$ are integration constants and $\omega\approx 0.5774$. 
Consequently, $\alpha\ll 1$ in the limit of $\xi\gg 1$.

\subsection{The Spherical Collapse Model}
While the independence of linear scalar perturbations on the field frame chosen is certainly 
reassuring, the relatively successful and simple description in the standard cosmological model of the nonlinear 
evolution of gravitationally bound objects down to virialization, and the key role they play 
in halo formation and evolution, compel us to demonstrate the equivalence of the 
spherical collapse model in the two frameworks, except 
for the effective `stiff matter' that dominates at very small field values 
which is important only at the would-be initial singularity. 
However, by the time BBN starts the effective stiff matter contribution is already negligible 
and so when matter overdensities start to collapse at the MD era, this contribution plays no 
role whatsoever. Final singularity is also irrelevant to the spherical 
collapse model because the latter is valid 
down to the virial radius anyway, and not beyond.

In the comoving frame, the potential 
of Eq. (9), $V(\phi)\equiv -K|\phi|^{2}+\lambda|\phi|^{4}$, 
has a nontrivial minimum for bound objects described by closed geometry $K>0$.
However, Eq. (9) was obtained by plugging the FRW curvature scalar, merely a solution 
of the classical field equations with an integration constant $K$, in the fundamental 
action, Eq. (8). The latter has no fundamental scale in it due to conformal invariance 
and consequently $K$ cannot induce $G$ and particle masses by 
a spontaneous symmetry breaking-like mechanism, even not in 
gravitationally bound objects. Inserting $K$ `by hand' into the fundamental action, 
as we did for motivational purposes in obtaining Eq. (9), would amount from our 
perspective to introducing a constant VEV into the SM action, or $G$ into the 
action describing the gravitational interaction.
Moreover, in the standard cosmological model it is required that the
background scale factor $a\propto\eta^{2}$ in the MD era agrees with the scale 
factor describing the evolving bound object $a\propto 1-\cos(c\eta)$ 
in the limit $\eta\rightarrow 0$, 
where here $c$ is a constant setting the allowed range of $\eta$. Similarly, 
continuity of $\phi$ and its first derivative during the linear evolution regime 
in the comoving frame forbids $\phi=constant$ in the interior of the gravitating body 
if $\phi$ evolves in the background.

As shown in Eqs. (15) \& (16), the background FRW equations are 
completely independent of the frame. In particular, they are 
equivalent in the case $K>0$, that corresponds to bound objects, and $a(\eta)$ in standard 
cosmology, is directly carried over to arbitrary frame, $a(\eta)\rightarrow\tilde{a}(\eta)$. 
The conserved quantity in an arbitrary frame is $\tilde{\rho}_{M}\propto\tilde{a}^{-(1+3w_{M})}$ 
rather than $\rho_{M}\propto a^{-3(1+w_{M})}$. Repeating the standard 
calculation of (generalized) NR overdensity evolution with respect to the background in the MD era, 
the classical result of the standard cosmological model is readily recovered. Therefore, both the 
exterior background and interior space inside gravitationally bound objects could be chosen to be, 
e.g. static, that only differ in the 
evolution histories of $\rho$ (the dilaton modulus) in the background and 
inside the bound object that control the evolution of overdensities, i.e. regions in 
space that gravitate over the background. The 
negative total energy of the evolving bound objects is carried by the dynamical scalar 
field, exactly as the scale factor $a$ that describes the dynamics of a gravitationally 
bound object carries negative total (kinetic) energy in the EF. 
Clearly, the scalar field is essentially the inverse of geometrized Newton gravitational constant, 
i.e. $l_{P}$, and inverse Compton wavelength $l_{c}$ up to a constant rescaling factor that 
is species-specific, namely, it varies, e.g., from electron to proton.

The standard parametric solution of the Friedmann equation in case of $K>0$ is 
$a(t)/a_{max}=(1-\cos\Theta)/2$ with $t/t_{max}=(\Theta-\sin\Theta)/\pi$.
The local conformal time coordinate $d\eta\equiv dt/a(t)$ is then related to $\Theta$ 
by $\eta=\frac{t_{max}}{a_{max}}\frac{2\Theta}{\pi}$. In a general frame, this 
cycloidal behavior of $a$ is replaced by an evolving $\tilde{a}$, 
i.e. $\tilde{a}(\eta)/\tilde{a}_{max}=[1-\cos(\eta/\eta_{m})]/2$, where 
$\eta_{m}\equiv 2t_{m}/(\pi a_{m})$. The smooth transition in the standard cosmological model of the 
scale factor from outside to inside the object as $a\propto t^{2/3}\propto\eta^{2}$ 
at the MD era is replaced in the comoving frame by a continuous transition of $\rho$, i.e. of the 
modulus scalar field proportional to particle masses and Newton constant, $\rho\propto\eta^{2}$. 
Spherical collapse is then understood as the local contraction of Compton wavelengths 
and Planck scales at a rate that is increasingly slower than that of 
the background until `overturn' is reached, followed by a gradual expansion of these 
fundamental length scales, which is normally interpreted as the collapse phase, until 
the structure is fully virialized. It should be mentioned that the model is parametrized 
by the parameter $\Theta$ which is defined over a finite range $[0,2\pi]$. Since in the 
model $\eta\propto\Theta$, then the spherical collapsing object must be singular at some initial  
conformal time $\eta=\eta_{ini}$. 

\section{Spherically Symmetric Static Vacuum Spacetime}
It is shown here that the spherically symmetric vacuum solution in conformal dilatonic 
gravity shares striking similarities with the corresponding solution obtained in Weyl's 
fourth-order, fully-conformal theory of gravitation. The solution obtained in the latter framework 
has been argued to provide a good fit to galactic rotation curves with no 
recourse to DM, e.g. [36, 37, 58, 111]. 
The freedom to locally select the conformal factor in the theory is 
exercised for achieving consistency with strong-lensing data [112]; 
the metrics for null and timelike geodesics differ by 
a coordinate transformation and a multiplicative conformal factor. 

An important distinction between vacuum solutions in fourth order 
Weyl gravity and dilatonic conformal gravity is 
that while in the former any {\it arbitrary} rescaling of the metric is a valid solution this 
conformal rescaling is not arbitrary in conformal dilatonic gravity: It 
must be accompanied by 
a corresponding rescaling of the scalar field, the latter functions as mass 
in our construction. As stressed earlier in this work, 
this implies that timelike geodesics are blind to conformal 
transformations since they are derived from $\mathcal{L}=(mc/\hbar)\int\frac{ds}{d\lambda}d\lambda$, 
the latter is proportional to the conformally invariant combination $\rho\sqrt{g_{\mu\nu}}$. 
Null geodesics on the other hand do not depend on $\rho$ but for the same reason they 
are indifferent to conformal transformations, e.g. [113]. Therefore, a given solution 
of the field equations in our conformal dilatonic gravity fully determines 
geodesics with no residual gauge-freedom.
In the following we derive the vacuum spherically symmetric static solution of 
conformal dilatonic gravity.

Consider the following spherically symmetric static line element
\begin{eqnarray}
ds^{2}=-B(r)d\eta^{2}+A(r)dr^{2}+r^{2}(d\theta^{2}+\sin^{2}\theta d\varphi^{2}).
\end{eqnarray}
The undetermined function $B(r)$, and with $A=B^{-1}$, will be fixed by the field equations.
To these we add the scalar field $\phi$, which stands for the space-dependent Newton `constant' 
$G\propto|\phi(r)|^{-2}$. The Einstein tensor components, and the curvature scalar $R$, 
constructed from this metric, are given by
\begin{eqnarray}
\frac{G_{\eta}^{\eta}}{B}&=&\frac{G_{r}^{r}}{B}=\frac{1}{r}\frac{B'}{B}+\frac{B-1}{Br^{2}}\nonumber\\
\frac{G_{\theta}^{\theta}}{B}&=&G_{\varphi}^{\varphi}=\frac{B''}{2B}
+\frac{1}{r}\frac{B'}{B}\nonumber\\
\frac{R}{B}&=&-\frac{B''}{B}-\frac{4}{r}\left(\frac{B'}{B}\right)-\frac{2(B-1)}{Br^{2}},
\end{eqnarray}
where all other components vanish, and in this section $f'\equiv\frac{df}{dr}$ for any function $f$.
Here, we consider the real field case, i.e. neglecting 
the effect of its phase dynamics, $\mathcal{G}_{\rho\rho}=-1$, $\mathcal{G}_{\theta\theta}=0$, 
and $F=\frac{\rho^{2}}{6}$.  
The $tt$-$rr$, $rr$, and $\theta\theta$ components of the generalized Einstein Eqs. (2), 
together with the scalar Eq. (3) applied to the modulus of the scalar field, result in
\begin{eqnarray}
2\left(\frac{\rho'}{\rho}\right)^{2}-\frac{\rho''}{\rho}&=&0\\
\left(\frac{1}{r}+\frac{\rho'}{\rho}\right)\frac{B'}{B}+\frac{B-1}{Br^{2}}&=&
-3\left(\frac{\rho'}{\rho}\right)^{2}-\frac{4}{r}\frac{\rho'}{\rho}\\
\frac{B''}{2B}+\left(\frac{1}{r}+\frac{2\rho'}{\rho}\right)\frac{B'}{B}&=&
-\frac{2\rho''}{\rho}+\left(\frac{\rho'}{\rho}\right)^{2}\nonumber\\
&-&\frac{2}{r}\frac{\rho'}{\rho}\\
\frac{\rho''}{\rho}+\left(\frac{B'}{B}+\frac{2}{r}\right)\frac{\rho'}{\rho}&+&
\frac{1}{6}\left(\frac{B''}{B}+\frac{4}{r}\frac{B'}{B}\right.\nonumber\\
&+&\left.\frac{2(B-1)}{Br^{2}}\right)=0.
\end{eqnarray}
From Eqs. (43)-(46), one easily recovers the canonical Schwarzschild metric $B=1-2r_{s}/r$, 
where $r_{s}$ is the Schwarzschild radius, if it is assumed that  $\rho=constant$. 

Next, we consider the case of non-constant $\rho$ and arrive at the general solution 
(up to conformal transformation of $g_{\mu\nu}$ and $\rho$)
\begin{eqnarray}
\rho&=&\frac{\rho_{0}}{1+\gamma r/(2-3\beta\gamma)}\nonumber\\
B&=&(1-3\beta\gamma)-\frac{\beta(2-3\beta\gamma)}{r}+\gamma r-\kappa r^{2} ,
\end{eqnarray}
where $\rho_{0}$, $\beta$ and $\gamma$ are integration constants. 
Interestingly, and perhaps not surprisingly, this metric solution coincides with 
the corresponding spherically static solution obtained in fourth order Weyl gravity [34, 35], 
except for the fact that the time coordinate appearing in Eq. (41) is conformal while 
the corresponding coordinate used in [34, 35] is cosmic.
As is assumed in [34, 35] we consider the limit $\beta\gamma\ll 1$ that guarantees 
the solution approaches the appropriate Schwarzschild limit at sufficiently small distances. 
Although we consider here a vacuum solution with a vanishing 
cosmological constant, $\kappa$ still functions as an effective cosmological constant, merely 
an integration constant appearing in the metric solution. Originally, the cosmological constant 
was indeed introduced to GR as an integration constant. Other proposals implying that the 
cosmological constant should be perceived as no more than an integration constant are, e.g. 
[114, 115]. Notably, conformal symmetry is asymptotically restored, and both the scalar field 
and metric in Eq. (46) scale by their canonical dimensions, i.e. $\rho\propto r^{-1}$ and 
$g_{\mu\nu}\propto r^{2}$.

In the standard cosmological model cosmological redshift does not 
depend on the local environment of the emitter 
but is rather an integrated effect due to space expansion between 
the emitter and the observer. Our argument 
in sections III.A-III.C, attributing cosmological redshift (fully or partially -- depending on the 
field frame) to variation of masses, applies only to the background cosmology. 
In order to explain cosmological redshift of light emitted from within gravitationally bound 
systems we utilize the local Weyl symmetry of the theory and rescale the fields in Eq. (47)
$\rho\rightarrow \rho/\tilde{a}(\eta)$, and $g_{\mu\nu}\rightarrow\tilde{a}^{2}g_{\mu\nu}$, 
where $\tilde{a}$ is the solution of Eqs. (15) \& (16). 
Consequently, space now expands in our vacuum solution, 
so do the fundamental length scales and therefore 
gravitationally bound systems look detached from 
the background expansion. As explained above, this conformal rescaling 
does not affect geodesics, both null and timelike. The evolution of length scales, 
e.g. the Rydberg `constant', within gravitationally-bound systems now conforms 
with that of the background and light is redshifted irrespective of whether 
the emitter is gravitationally bound or not.

As noted in [34], the Weyl tensor components 
associated with this solution are proportional to $\beta(2-3\beta\gamma+\gamma r)/r$, so that
this spacetime is conformally flat in case $\beta=0$, i.e. the underling conformal invariance of 
the theory, Eq. (8), is broken by the introduction of the dimensional integration constant, 
$\beta$, i.e. the Schwarzschild radius. 
In the limit $r\gg\beta$ and $r\gg\gamma^{-1}$, the Weyl tensor is proportional to $\beta\gamma$, 
which by assumption is negligibly small compared to unity.
Although our metric solution is identical to that 
obtained from Weyl gravity, it exhibits a nontrivially varying $\rho$. 
This field is absent from vacuum solutions of Weyl gravity -- it appears only in the 
matter sector of the fundamental action, e.g. [62]. It is clear from the solution that 
although locally $\rho$ decreases away from the sources of the static gravitational 
field, it is practically constant insofar as $r\ll 2/\gamma$. 

It has been found from fitting rotation curves [36, 37] 
that $\gamma=\gamma_{0}+\left(\frac{M}{M_{\odot}}\right)\gamma_{\star}$ 
where $M$ is the source mass, 
$\gamma_{0}=3.05\times 10^{-30} cm^{-1}$, and 
$\gamma_{\star}=5.42\times 10^{-41} cm^{-1}$ are `universal' 
constants over the relevant galactic mass range.
This sample has been further extended and the fit has been applied to dwarf galaxies 
as well [116]. Indeed, at larger distances masses are space-dependent and 
trajectories deviate from their standard geodesic motion [25]. 

The standard cosmological model is able to explain both galactic rotation curves 
and strong-lensing data with the same amount of CDM 
provided that its density profile has a specific form. 
Interestingly, inference from strong-lensing data and galactic 
rotation curves based on Weyl gravity implies that the corresponding 
$\gamma$ values have opposite signs. 
The fact that $\gamma$ has to be negative and positive for lensing and galactic rotation 
curves, respectively, was pointed out in [117].
This effective sign flip of $\gamma$ finds a natural explanation, 
much like in the equivalent solution obtained from fourth order Weyl gravity [112];
the underlying conformal symmetry of the theory has been utilized, in combination 
with an appropriate coordinate transformation to reverse the sign of $\gamma$ in going 
from null to timelike geodesics. 
Here we follow the same methodology, but since our solution 
is more constrained in the sense that $\phi$ cannot be arbitrarily chosen as 
in [112] (but is rather determined self-consistently 
with $g_{\mu\nu}$ from the field equations), our coordinate transformation is different. 
We find that in order to reverse the (minus) sign of the linear $\gamma r$ term in $B(r)$ 
of Eq. (47), in transforming to the effective metric for timelike geodesics, the following 
coordinate transformation $t'=t/\rho$ and $r'=\rho^{3}r$ has to be invoked.
This establishes consistency between galactic rotation curves and 
strong lensing data.

On solar system scales, and considering the linear potential term as an effective 
scale-dependent correction to the Schwarzschild radius, 
i.e. $GM$, and given current precision $\delta G/G=O(10^{-4})$, e.g. [118], 
one obtains the empirical upper bound $\gamma\lesssim 10^{-4}\beta/r^{2}$. 
On solar system scales GR was found to be a very good 
fit out to at least $500 AU$ from the sun, where a few anomalies [119]
have been observed that prompted the `planet nine' hypothesis. 
This scale translates to $\gamma\lesssim 
10^{-32} cm^{-1}$ for an Earth mass object. 

As to the Newtonian limit of the model, 
$2\varphi_{grav}=-2\beta/r+2\gamma r-\kappa r^{2}\rightarrow -2\beta/r$, the 
gravitational force is always attractive at sufficiently small distances, i.e. the sign 
of $\beta$ (though not its strength) is universal, much like the sign of $r_{s}$ is 
always positive in GR. 
The postulated universality of positive $G$ and particle masses then lead to GR. 
Here we similarly postulate that the ratio $r_{s}/l_{C}=GMm/(2c\hbar)$ is universally 
positive. Both postulates are equivalent to 
stability requirements on the fundamental action.

It is unclear, as of yet, whether a `bullet'-like cluster 
can be explained by our model with no recourse to CDM, as dynamics on galactic scales does. 
It has been claimed that whereas modified Newtonian dynamics (MOND) removes the need for 
CDM on galactic scales, e.g. [120, 121], a `bullet' cluster 
does pose a challenge to MOND, requiring a residual amount of CDM, 
though only at a level comparable to baryonic matter, and not 5-6 times larger, 
as in the standard cosmological model. 
It is not inconceivable that an entirely DM-free 
viable model based on conformal dilatonic gravity 
will eventually be formulated. In fact, there are 
counter-examples to the often-made claim that bullet-like clusters cannot be explained without 
CDM; e.g., in the context of theories of modified gravity, e.g [122].

If clumped CDM is not required on galactic scales but is nevertheless required on 
cluster scales, especially in bullet-like systems, it implies that CDM particles are ultralight with 
typical de Broglie wavelength $\lambda_{dB}=\hbar/(m_{DM}v)$, longer than galactic scales, 
but shorter than typical cluster scales, where here $v$ stands for the particle velocity. 
Since our solution was shown to obviate CDM on $\approx 20$ kpc or smaller, but may fail 
to explain bullet-cluster-like phenomena on $0.1$ Mpc 
scales or larger, then since the shortest de Broglie wavelength is the Compton wavelength, we 
obtain a lower limit on the CDM particle mass of 
$m_{DM}\gtrsim 10^{-23} eV/c^{2}$, implying that this species might provide the leading 
contribution to the total entropy in a Hubble volume (in case that the weak 
inequality becomes an equality), $\mathcal{S}=O(10^{111})$, far in excess of the largest 
known entropy contributor -- supermassive black holes, e.g. [123]. 

We recall that the cosmological model, discussed in section III, underscored the privileged 
role played by conformal- over cosmic-time. It is encouraging 
that a similar conclusion could be reached based on the spherically symmetric vacuum solution. 
As shown in [34] the line element Eq. (41), with $B(r)$ given by Eq. (47) and 
on scales much larger than $\sqrt{\beta/\gamma}$ and $(\beta/\kappa)^{1/3}$,
could be transformed to a line element conformally-related to that of a 
FRW spacetime by carrying out the following coordinate transformations
\begin{eqnarray}
\xi &\equiv&\frac{4r}{2(1+\gamma r-\kappa r^{2})^{1/2}+2+\gamma r}\nonumber\\
t&\equiv&\int a(\eta)d\eta.
\end{eqnarray}
The transformed line element then reads
\begin{eqnarray}
ds^{2}=\Omega^{2}\left(d\eta^{2}-\frac{d\xi^{2}}{1-K\xi^{2}}-\xi^{2}(d\theta^{2}+\sin^{2}\theta d\varphi^{2})\right) ,
\end{eqnarray}
where $\Omega\equiv\frac{1-\xi^{2}(\gamma^{2}/16+\kappa/4)}{(1-\gamma\xi/4)^{2}+\kappa\xi^{2}/4}$ 
and $K\equiv -\kappa/4-\gamma^{2}/16$, which implies that for an effective positive cosmological constant, 
i.e. $\kappa>0$, the modified Schwarzschild-de Sitter solution is conformally related to an open FRW spacetime, 
consistent with our choice 
$K<0$ in section III. Comparison of Eq. (49) with the standard form of the FRW metric reveals that the 
time coordinate appearing in Eq. (41) should indeed be the conformal time. Therefore, the fundamental 
conformal symmetry of conformal dilatonic gravity and the smooth extrapolation of the Schwarzschild-de Sitter 
metric to the embedding FRW spacetime, reveal the genuine nature of 
the (conformal) time coordinate, providing further evidence that 
what is normally considered `cosmic time' is an artifact of our conventional unit system in which 
particle masses are fixed. 

Finally, the constancy of dimensional integration {\it constant} $\beta$, essentially 
the Schwarzschild radius, may seem at first glance at odds with the scaling 
of $GM/c^{2}\propto\tilde{a}^{-1}$ at the background. However, as explained 
above in the context of redshift from a gravitationally bound emitter, a conformal 
transformation could be applied to the static metric to look expanding, 
$g_{\mu\nu}\rightarrow\tilde{a}^{2}g_{\mu\nu}$, i.e. effectively $r\rightarrow\tilde{a}r$, 
which is equivalent to $r_{S}\rightarrow r_{S}/\tilde{a}$. In these `expanding' coordinates 
the integration constants $\beta$ and $\kappa$ (i.e. the cosmological constant) 
are promoted to time-dependent functions and are now on par with the corresponding 
background quantities. 

\section{Summary}
While it is unquestionable that the standard cosmological model has been very 
successful in phenomenologically describing a wide spectrum of observations, it is fair to 
say that it still lacks a microphysical description of a few key ingredients, 
e.g. CDM and DE. In addition, {\it direct} 
spectral information on the CMB is unavailable (due to opacity) in the pre-recombination 
era, $z\gtrsim 1100$. 
From the cosmic abundance of light elements, 
the BBN at redshifts $O(10^{9})$ could be indirectly probed. Earlier on, at 
energy scales of $O(200)$ MeV and $O(100)$ GeV, the QCD and electroweak phase transition 
had presumably occurred, although their (indeed week) hallmark signatures in the CMB and LSS 
have not been found. In addition, inflation, a cornerstone in the standard cosmological model, 
is clearly beyond the realm of well-established physics; its detection via 
the B-mode polarization it induces in the CMB could be achieved only 
if it took place at energy scales $\sim 13$ orders of 
magnitude larger than achievable at present.

Assuming that conformal invariance is an exact symmetry of the four fundamental 
interactions, it was argued here that we are led to abandon primordial phase transitions, 
galactic DM, and perhaps even cosmic inflation if the latter is to 
generate metric perturbations from a real (rather than complex) 
inflaton field. Indeed, local scale invariance, i.e. the freedom 
to choose our fundamental length scales, is a powerful 
symmetry principle that severely constrains cosmological 
scenarios and types of interactions, even on microphysical scales.

By and large, any viable formulation of the fundamental 
interactions should {\it ideally} be independent of 
our system of units, in a similar fashion to its independence of the 
choice of a coordinate system. 
Much as GR is formulated in a covariant fashion, with only the 
symmetries of physical systems as a basis to favor one coordinate system over others, 
so should a more advanced theory of gravitation and the other fundamental interactions 
be preferably formulated in a fashion that has no {\it a priori} preference to one unit system or 
another. The latter is determined by the manifest symmetries of the (physical) 
system configuration.

Yet, our theories 
of fundamental interactions are non-conformal. For example, the Higgs VEV provides 
a typical (conformal) symmetry-breaking scale at the action level that 
determines $l_{C}$, the Compton wavelengths of massive particles. Similarly, GR 
contains two additional typical length scales -- $l_{P}$ and $l_{\Lambda}$, 
the Planck length, and the length scale associated with the 
cosmological constant, respectively. 
In other words, in standard field theory these 
characteristic scales are kinematic input parameters, rather than 
dynamically generated via, e.g., spontaneous symmetry breaking mechanism.

What experimental evidence provides the basis 
for this conventional choice of (constant) units? 
Could experiments guide us,  
even in principle, about the `appropriate' system of units to be used (i.e. 
the units system allegedly realized in nature)? 
As said, observations can only hint towards the fundamental 
symmetries of the physical systems under consideration, thereby preferring one unit system 
over others, akin to the 
preference of one coordinate system over others. For example, the underlying global 
$U(1)$ symmetry of the proposed conformal theory (and the cosmological model derived from it), 
i.e. the dependence of the non-minimal coupling to gravity, 
essentially Newton gravitational constant, and potential terms only on the modulus of the 
dilaton field, reflects the observed isotropy on {\it cosmological} scales.

Much like `unnatural' choices of coordinate systems 
can lead to entirely contrived and serendipitously 
misleading interpretations or even result in incorrect inferences, 
the same might happen if an unnatural system of units is used. 
For example, spacetime variation of fundamental dimensional 
quantities (described as classical fields 
in our work) in a given system of units generally carries energy and momentum, and 
arbitrarily fixing these units based on prejudice -- or worse, `convenience' -- 
hides this contribution to the total energy-momentum budget. This missing energy
must then be compensated for by other dynamical fields (or worse by parameters that conceptually 
provide only a poor substitute to these dynamical fields) to explain observations.
For example, as we argued in this work, space expansion might just be a (perhaps convenient) 
representation of varying physical `constants' in static space with constant relative strength 
of the fundamental interactions. In the latter case there is no initial curvature singularity.
Although this singularity is now `absorbed' by singular scalar fields, i.e. by the 
varying `constants', it should be admitted that singularity in scalar and gauge fields 
are more easily cured in the quantization process than singular metrics are; 
GR is a non-renormalizable theory, again -- due to the (fixed) Planck scale, i.e. the 
fact that the coupling {\it constant} $G$ has positive mass dimension. 
In comparison, conformal dilatonic gravity contains no such dimensional 
constant at the action level.

In addition, these units (scalar fields) may fluctuate, thereby inducing primordial 
metric perturbations. For example, we showed in this work that the cosmic scale factor might 
be viewed as the modulus 
of the complex dilaton field. Fixing the phase of the dilaton (as is {\it effectively} done in 
the standard cosmological model) ignores its possible perturbations as well, the latter are 
described by nearly scale-invariant perturbations. In the standard cosmological model we 
then obtain the desired perturbations from the fluctuations of another field -- the inflaton. 

On the other hand, allowing 
for particle masses, $G$, and $\Lambda$ to vary, generally results in energy-momentum 
non-conservation.
We stress that while energy-momentum conservation usually breaks down 
under local conformal (Weyl) transformation for non-traceless perfect fluids, thereby 
resulting in non-geodesic timelike trajectories, this is not necessarily in 
conflict with observations. On the contrary, one might even argue that forcing energy-momentum 
conservation on galactic scales led to invoking 
CDM in order to explain the observed anomalous rotation curves and strong lensing data, 
and prompted us also to postulate the existence of the cosmological constant to explain the 
recent cosmic acceleration phase. While both DE and CDM have been rather successful in 
{\it parametrically} fitting observations to GR predictions, the microphysics 
of CDM and DE is still a puzzle. 

Since it is conventionally assumed that $l_{P}$, 
$l_{C}$, and $l_{\Lambda}$ are fixed, the observed cosmological redshift can only be 
explained by space expansion. However, `expansion' is a relative term. 
We have shown that the 
Friedmann equations are invariant to simultaneous rescaling of particle masses and the FRW metric, 
and that cosmological redshift is generally accounted for by a combination of the evolving 
metric and scalar fields, i.e. evolving particle masses. 
The very notion of space expansion can be attributed to 
forcing energy-momentum conservation on our fundamental theory of gravity, GR. 
Relaxing this -- essentially no more than units convention -- results in point particles 
following non-geodesic paths. In the cosmological context of homogeneous and isotropic 
space this is manifested in modifications in the temporal component, i.e. leads to the systematic 
redshift observed on cosmological scales, due both to stretching of photon wavelength 
along its path 
to our telescopes, but also due to a monotonic variation of the Rydberg constant over 
cosmological time scales. In the special case of the `comoving frame', space is fixed 
(i.e. the metric is static), while the fundamental length scales 
contract. From this perspective, and on a more fundamental level, 
the apparent space expansion on cosmological scales might 
be viewed as a manifestation in, e.g., the comoving frame 
of the time-dependent VEV of the Higgs field. 
This field is identified up to a constant rescaling 
with the dynamical scale factor $a$ of the standard cosmological model in the 
post-bounce phase of the non-singular cosmological model advocated here. 
This description in the comoving frame provides 
a Machian version of the standard cosmological model: 
In this specific frame energy densities 
regulate the contraction rate of the fundamental particle masses, i.e. 
particle masses are determined locally by the average energy density 
in this static spacetime. This neat picture is somewhat obscured in a 
general field frame, and is completely lost in the standard cosmological 
model that relies on GR and its underlying unit system conventions.

In the standard cosmological model the fundamental time coordinate is `cosmic time', and 
conformal time is only defined for convenience using the scale factor -- the latter satisfies 
dynamical equations, e.g. the Friedmann equation in a GR-based cosmology. From the way 
conformal time is defined, $d\eta\equiv dt/a(t)$, and the fact that $a(t)$ is monotonically 
increasing in the standard cosmological model, 
it is clear that the `cosmic clock' was effectively ticking slower for photons 
in the past than it did for massive particles. Since in the EF particle masses, and thus 
the Rydberg constant, are fixed this implies that light emission from a distant object must 
have effectively redshifted. In contrast, our model highlights the status 
of conformal time $\eta$ as the fundamental time coordinate on all scales, 
and cosmic time plays no fundamental role; the `cosmological clock' is 
universal -- it ticks at the same rate for both matter and radiation 
[at the `expense' of allowing masses to evolve as $a(t)$ would in the 
standard cosmological model]. 
The observed cosmological redshift must therefore be explained in this frame 
by varying masses, or more concretely -- varying Rydberg `constant'. 

While in the present work all fundamental physical 
quantities such as the Compton wavelenths of particles and the Planck 
length are promoted to dynamical scalar fields dimensionless 
polynomial ratios thereof are fixed, by definition, in agreement with 
observations. Since for the entire cosmic history particle masses and the 
Planck mass have the same dynamics, the dimensionless 
gravitational coupling $\alpha_{g}\equiv Gm_{1}m_{2}$ (between two 
masses $m_{1}$ \& $m_{2}$), and with it the relative strength of 
electromagnetic-to-gravitational interactions $\alpha_{e}/\alpha_{g}$, 
are kept constant, and the sequence of cosmological epochs is in principle 
as in the standard cosmological model. Our very early universe scenario is 
significantly different from the standard cosmological one only because we 
effectively introduced additional degree of freedom -- the dilaton phase. 
This is further discussed below.

Our reformulation of the background field equations 
and linear perturbations in arbitrary field frames demonstrates that 
cosmic evolution, including the evolution of gravitationally bound 
objects, such as halos, is frame-independent. The dilaton phase 
and its perturbations do not appear in the standard cosmological model; the FRW metric 
in the latter is determined by a single scale factor -- essentially the 
modulus of our complex dilaton field. 

An early inflationary phase, i.e. linear increase in (conformal) time of the 
fundamental length scales, certainly remains a viable solution in  
an arbitrary frame description to the classical horizon, flatness, 
and magnetic monopole problems. However, perturbations of 
the (real) inflaton field cannot explain primordial metric 
perturbations in the conformal dilaton framework because the latter does not 
allow modulus field perturbations. 

A tantalizing alternative explored in this 
work is a pseudo-conformal deflationary evolution phase 
followed by a curvature-, matter-, and `stiff matter'-dominated phases. 
During the pseudo-conformal 
deflation epoch, which could be set to arbitrary long conformal time, gaussian adiabatic 
scalar perturbations characterized by a red-tilted spectrum, which are 
sourced by the (quantum) fluctuating phase of the complex dilaton field, 
are generated.
As in inflation, the observed gaussianity of scalar perturbations is explained 
by the correspondence of the (quantum) vacuum state to the ground state of an harmonic oscillator.
It is then maintained due to a very small self-coupling of the dilaton phase fluctuations
during the conformal epoch, combined with the fact that the leading order modulus perturbation 
vanishes and therefore cannot couple to the transversal modes, at leading order at least, 
and spoil gaussianity, as occurs in similar scenarios.
The adiabatic nature of density perturbations is explained as a consistency requirement 
for linear perturbations of the metric, matter, and the dilaton, which is unique to conformal 
dilatonic gravity. The latter is absent 
in GR and adiabacity has to be achieved by invoking inflation.
Should certain level of non-adiabatic perturbations modes 
be detected in the future, our conformal model is immediately falsified, unlike 
in the case of standard cosmology inflation where isocurvature perturbations 
could be explained by multi-field scenarios.

The scalar field does not `slow-roll' 
along its potential (as in the standard inflationary scenario) 
but rather its kinetic energy is comparable to its potential 
energy. It therefore does not suffer from the fine-tuning of the inflaton 
potential generically required in the standard cosmological model.
Again, from the present work perspective `slow-roll' is an artifact of the 
standard units convention, in which all scalar fields (e.g. particle masses, 
$G$, $\Lambda$, the inflaton itself, etc.) are effectively set to be constants. 

The cosmological model is non-singular due to an interplay between the negative energy density 
associated with the kinetic phase term and the positive energy density associated with 
relativistic matter (and possibly with a nonvanishing negative spatial curvature).
It should be stressed that the bounce is classical and does not rely 
on quantum gravitational effects. 
This bounce is followed by an effectively stiff matter phase and the conventional 
expansion RD, MD, and DE epochs familiar from the standard cosmological model.
The relative amount of effective stiff matter has to be sufficiently small as 
not to conflict with BBN constraints, as well as not to deviate too much from 
the standard history of CMB thermalization in the very early universe.

We have shown that linear perturbations generated during the pseudo-conformal phase 
generically survive the bounce and can be kept sufficiently small as to not undermine 
the underlying homogeneity and isotropy of the cosmological model in the post-bounce phase. 
Matter is not created in the (non-singular and adiabatic) cosmological scenario 
layed out here, nor is it destroyed. The cosmological scenario from the BBN era 
(and possibly even earlier) on is exactly as in the standard cosmological model.

Both conformal and cosmic times in this scenario are in fact past-unbounded 
and this formally implies that there is no `horizon problem' associated with it -- 
not for radiation, and not even for, e.g. light (but still massive) neutrinos. 
Specifically, the pre-bounce starts from a very large 
$\tilde{a}$, much larger than its present value (and in principle infinite) and therefore 
the causal horizon is much larger than would be naively expected from a monotonically 
expanding space. Likewise, the `flatness problem' afflicting the hot big bang scenario 
stems from the slower decay of the energy density associated with curvature as compared to 
that of matter in a {\it monotonically} expanding universe. In our bouncing scenario the 
situation is reversed in the pre-bounce phase; starting at $\eta=-\infty$ one typically 
expects to find that the energy density in the form of matter largely exceeds that of 
curvature at any {\it finite} time in the pre-bounce phase. 
Since this adiabatic model is very nearly symmetric 
in $\tilde{a}$ in the pre- and post-bounce phases (barring entropy 
processing effects at around the RD phases), one generically expects the universe to 
look spatially flat at any finite (conformal) time after the would-be singularity 
(actually a non-singular bounce). From that perspective flatness is an 
attractor-, rather than an unstable-point that requires fine-tuning. The `monopole' and 
`relic defects' problems do not arise (in the proposed scenario) in the first place because 
there are no primordial phase transitions associated with conformally coupled scalar fields 
-- their thermal fluctuations vanish.

The underlying isotropy on cosmological scales is manifested by 
the {\it global} $U(1)$ symmetry of the model, i.e. the 
dependence of both the potential- and non-minimal 
coupling to gravity- terms only on the field modulus. 
Therefore, exactly opposite to the standard inflationary case, 
perturbation of the transversal mode necessarily comes from the kinetic term.
No analog of this pseudo-Nambu-Goldstone-like boson origin for gravitational potential 
perturbations could be found for either PGW or vector perturbations. Therefore 
any detection of such signatures of cosmological origin 
-- e.g. via B-mode detection of the CMB polarization -- would likely require 
an alternative mechanism for generating perturbations. 
Unlike in the SM of particle physics where the Goldstone boson, 
the phase of the Higgs field, is gauged away by transformation of gauge fields, 
the dilaton phase is decoupled from any gauge (or any other) field and is therefore 
a genuine degree of freedom of the model. For exactly the same reason it has 
not been detected.

Conformal dilatonic gravity admits spherically symmetric vacuum solutions for 
a modified Schwarzschild-de Sitter spacetime augmented by a linear potential term 
that includes three integration constants, thereby introducing three length scales, 
one of which -- the Hubble scale -- is universal. The other two are case-specific. 
When applied to galactic (or larger) scales, 
this approach results in significant departures from standard interpretations of observations 
on super-galactic scales. This pertains, in particular, to 
our understanding of the nature of cosmological redshift, DM, and DE.
This implies that CDM may not be required on galactic 
and sub-galactic scales, but may still be needed for a proper phenomenological 
description of phenomena on galaxy cluster scales and larger, 
e.g. `bullet'-like systems. This fact alone already sets a lower 
bound, $m_{DM}\gtrsim 10^{-23} eV/c^{2}$, on the mass of CDM particles. 

\section*{Acknowledgments}
The author is indebted to Yoel Rephaeli for numerous constructive, critical, and 
thought-provoking discussions which were invaluable for this work.

\appendix

\section{Background and Linear Perturbation Equations}
Here we provide the self-contained system of background equations, as well as 
linear perturbation equations over the FRW. The equations of [38, 39] are here 
recasted in conformal time coordinates

\subsection{Background Equations}
The Friedmann, Raychaudhuri, and scalar field equations are, respectively
\begin{eqnarray}
\mathcal{H}^{2}+K&=&\frac{a^{2}(\rho_{M}+V)}{3F}+\frac{g_{IJ}\phi'^{I}\phi'^{J}}{6F}\nonumber\\
&-&\mathcal{H}\frac{F'}{F}\\
\mathcal{H}'-\mathcal{H}^{2}-K&=&-\frac{a^{2}\rho_{M}(1+w_{M})}{2F}-\frac{g_{IJ}\phi'^{I}\phi'^{J}}{2F}\nonumber\\
&-&\frac{F''}{2F}+\mathcal{H}\frac{F'}{F}\\
\phi''^{I}+2\mathcal{H}\phi'^{I}&+&\Gamma_{JK}^{I}\phi'^{J}\phi'^{K}+a^{2}(V^{,I}+\rho^{,I}_{M})\nonumber\\
&-&\frac{F^{,I}}{2}(a^{2}R)=0
\end{eqnarray}
where the connection is $\Gamma^{I}_{JK}\equiv\frac{1}{2}g^{IL}(g_{LJ,K}+g_{LK,J}-g_{JK,L})$ 
and the curvature scalar $R$ is given by $a^{2}R/6=\mathcal{H}^{2}+ \mathcal{H}'+K$. 
From the above relations the generalized continuity equation is obtained
\begin{eqnarray}
\rho'_{M}+3(1+w_{M})\mathcal{H}\rho_{M}=\rho_{M,I}\phi'^{I}.
\end{eqnarray}

\subsection{Linear Order Perturbation Equations}
The perturbation equations for the linear gravitational 
potentials $\varphi$, $\alpha$, the velocity perturbation $v$, the shear $\chi$, 
the expansion $\kappa$, the spatial curvature perturbation $\delta R$, the perturbations 
of energy density and pressure $\delta\rho_{M}$ and $\delta P_{M}$, and perturbation of the scalar 
field $\delta\phi$, are given in [38, 39] in the presence of anisotropic stress $\pi^{(s)}$. 
The latter is sourced mainly 
by neutrino free-streaming, and to a lesser extent by CMB photons freely streaming 
since recombination. The evolution of the anisotropic stress is calculated 
self-consistently with the evolution of metric and density perturbations using the 
collisionless Boltzmann equation and a Boltzmann equation with a 
Thomson collision term for neutrinos and photons, respectively.

The ADM energy (time-time), momentum (time-space), and propagation (space-space) 
components respectively read
\begin{eqnarray}
&-&(k^{2}-3K)\varphi+\left(\mathcal{H}+\frac{F'}{2F}\right)(a\kappa)-\frac{1}{2F}(g_{IJ}\phi'^{I}\phi'^{J}-3\mathcal{H}F')\alpha\nonumber\\
&=&-\frac{(a^{2}\rho_{M})\delta_{\rho_{M}}}{2F}-\frac{g_{IJ}\phi'^{I}\delta\phi'^{J}}{2F}\nonumber\\
&-&\left[g_{IJ,K}\phi'^{I}\phi'^{J}-a^{2}(F_{K}R-2V_{K})\right]\frac{\delta\phi^{K}}{4F}\nonumber\\
&+&\frac{3\mathcal{H}}{2}\frac{\delta F'}{F}+(k^{2}-3\mathcal{H}')\frac{\delta F}{2F}\\
&&(a\kappa)-(k^{2}-3K)\left(\frac{\chi}{a}\right)+\frac{3}{2}\frac{F'}{F}\alpha\nonumber\\
&=&\frac{3}{2F}\left[a^{2}\rho_{M}(1+w_{m})\left(\frac{v}{k}\right)+g_{IJ}\phi'^{I}\delta\phi^{J}+\delta F'-\mathcal{H}\delta F\right]\\
&&\left(\frac{\chi}{a}\right)'+\left(2\mathcal{H}+\frac{F'}{F}\right)\left(\frac{\chi}{a}\right)-\alpha-\varphi=
\frac{a^{2}\pi^{(s)}}{Fk^{2}}+\frac{\delta F}{F}
\end{eqnarray}
where we defined $\delta_{\rho_{M}}\equiv\delta\rho_{M}/\rho_{M}$ and 
$\delta_{P_{M}}\equiv\delta P_{M}/\rho_{M}$, and the `expansion' perturbation parameter
\begin{eqnarray}
a\kappa\equiv 3\mathcal{H}\alpha+k^{2}\left(\frac{\chi}{a}\right)-3\varphi'.
\end{eqnarray}
The Perturbed Raychaudhuri, scalar fields, and trace of the Einstein tensor, result in the following equations
\begin{eqnarray}
&&(a\kappa)'+\left(\mathcal{H}+\frac{F'}{2F}\right)(a\kappa)+\frac{3}{2}\frac{F'}{F}\alpha'\nonumber\\
&+&\left(3\mathcal{H}'-3\mathcal{H}^{2}+3\frac{F''}{F}-\frac{3\mathcal{H}}{2}\frac{F'}{F}
+\frac{2g_{IJ}\phi'^{I}\phi'^{J}}{F}-k^{2}\right)\alpha\nonumber\\
&=&\frac{a^{2}\rho_{M}(\delta_{\rho_{M}}+3\delta_{P_{M}})}{2F}+\frac{2g_{IJ}\phi'^{I}\delta\phi'^{J}}{F}\nonumber\\
&+&\frac{g_{IJ,K}\phi'^{I}\phi'^{J}\delta\phi^{K}}{F}+\frac{a^{2}(F_{K}R-2V_{K})\delta\phi^{K}}{2F}\nonumber\\
&+&\frac{3\delta F''}{2F}+(k^{2}-6K-6\mathcal{H}^{2})\frac{\delta F}{2F}\\
&&\delta\phi''^{I}+2\mathcal{H}\delta\phi'^{I}+2\Gamma^{I}_{JK}\phi'^{J}\delta\phi'^{K}+k^{2}\delta\phi^{I}\nonumber\\
&+&\left[a^{2}V^{;I}_{L}-\frac{F^{;I}_{L}}{2}(a^{2}R)+\Gamma^{I}_{JK,L}\phi'^{J}\phi'^{K}\right]\delta\phi^{L}\nonumber\\
&=&\phi'^{I}(a\kappa+\alpha')+(2\phi''^{I}+\mathcal{H}\phi'^{I}+2\Gamma^{I}_{JK}\phi'^{J}\phi'^{K})\alpha\nonumber\\
&+&\frac{F^{;I}}{2}a^{2}\delta R-a^{2}\delta\rho_{M}^{I}\\
&&\delta F''+2\mathcal{H}\delta F'+\left(k^{2}-\frac{a^{2}R}{3}\right)\delta F+\frac{2}{3}g_{IJ}\phi'^{I}\delta\phi'^{J}\nonumber\\
&+&\frac{1}{3}\left[g_{IJ,K}\phi'^{I}\phi'^{J}+2(F_{K}a^{2}R-2a^{2}V_{K})\right]\delta\phi^{K}\nonumber\\
&=&\frac{a^{2}\rho_{M}(\delta_{\rho_{M}}-3\delta_{P_{M}})}{3}+F'(a\kappa+\alpha')\nonumber\\
&+&\left(\frac{2}{3}g_{IJ}\phi'^{I}\phi'^{J}+2F''+\mathcal{H}F'\right)\alpha-\frac{Fa^{2}\delta R}{3}
\end{eqnarray}
where the perturbed curvature scalar is given by
\begin{eqnarray}
\frac{a^{2}\delta R}{2}&=&-(a\kappa)'-3\mathcal{H}(a\kappa)+(k^{2}-3\mathcal{H}'+3\mathcal{H}^{2})\alpha\nonumber\\
&+&2(k^{2}-3K)\varphi.
\end{eqnarray}
The system is closed by the generalized continuity and Euler equations
\begin{eqnarray}
&&\delta'_{\rho_{M}}+\left(3\mathcal{H}+\frac{\rho'_{M}}{\rho_{M}}\right)\delta_{\rho_{M}}+3\mathcal{H}\delta_{P_{M}}
-\frac{\rho_{M,I}}{\rho_{M}}\delta\phi'^{I}\nonumber\\
&=&(1+w_{m})(a\kappa-3\mathcal{H}\alpha-kv)+\phi'^{I}\frac{\delta\rho_{M,I}}{\rho_{M}}\\
&&\frac{[a^{4}\rho_{M}(1+w_{m})v]'}{a^{4}\rho_{M}(1+w_{m})k}=\alpha+\frac{\delta_{P_{M}}}{1+w_{m}}\nonumber\\
&-&\frac{2}{3}\frac{(1-3K/k^{2})\pi^{(s)}}{\rho_{M}(1+w_{m})}+\frac{\rho_{M,I}\delta\phi^{I}}{\rho_{M}(1+w_{m})}.
\end{eqnarray}

\end{document}